\documentclass[runningheads]{llncs}

\usepackage{rotating}
\usepackage{graphics}
\usepackage{latexsym}
\usepackage[dvips]{epsfig}
\usepackage{amssymb}
\usepackage{graphicx}

\usepackage{url}

\def\boxit#1{\vbox{\hrule\hbox{\vrule\kern4pt
  \vbox{\kern1pt#1\kern1pt}
\kern2pt\vrule}\hrule}}

\def\qed{\rule{1.5mm}{3mm}}

\newcommand\nc{\newcommand}

\nc{\crl}[2]{\begin{corollary}\label{crl:#1} #2 \end{corollary}}
\nc{\dfn}[2]{\begin{definition}\label{def:#1} #2 \end{definition}}
\nc{\lem}[2]{\begin{lemma}\label{lem:#1} #2 \end{lemma}}
\nc{\prp}[2]{\begin{proposition}\label{prp:#1} #2
\end{proposition}}
\nc{\thm}[2]{\begin{theorem}\label{thm:#1} #2\end{theorem}}
\nc{\fac}[2]{\begin{lemma}\label{fact:#1} #2 \end{lemma}}

\nc{\eqn}[2]{\begin{eqnarray}\label{eqn:#1} #2 \end{eqnarray}}

\nc{\fig}[4]{\begin{figure}[h]
\begin{center}
\includegraphics[width=#2\textwidth]{#4}
\end{center}
\caption{#3}\label{fig:#1}
\end{figure}}

\nc{\tbl}[3]{\begin{table}[hbt] #3 \caption{#2} \label{tab:#1}
\end{table}}

\nc{\refc}[1]{Corollary~\ref{crl:#1}}
\nc{\refd}[1]{Definition~\ref{def:#1}}
\nc{\reff}[1]{Figure~\ref{fig:#1}}
\nc{\refl}[1]{Lemma~\ref{lem:#1}}
\nc{\refp}[1]{Proposition~\ref{prp:#1}}
\nc{\reft}[1]{Theorem~\ref{thm:#1}} \nc{\refe}[1]{(\ref{eqn:#1})}
\nc{\reftb}[1]{Table~\ref{tab:#1}}

\nc{\reffc}[1]{Fact~\ref{fact:#1}}

\nc{\pf}[1]{ \noindent \emph{Proof.} #1 \hfill \qed\par}

\long\def\invis#1{}

\begin{document}



\title{Exact Algorithms for Maximum Independent Set}

\author{Mingyu Xiao\inst{1}\thanks{Supported by NFSC of China under the Grant
61370071 and Fundamental Research Funds for the Central Universities under the Grant ZYGX2012J069.} \and
Hiroshi Nagamochi\inst{2}}

 \institute{
 School of Computer Science and Engineering,
University of Electronic Science and Technology of China, China,
 \email{myxiao@gmail.com}
 \and
 Department of Applied Mathematics and Physics,
  Graduate School of Informatics, Kyoto University, Japan,
 \email{nag@amp.i.kyoto-u.ac.jp}}
%



%
%

\toctitle{Maximum Independent Set} \tocauthor{}
\maketitle

\vspace{-4mm}
\begin{abstract}
We show that the maximum independent set problem (MIS)  on an $n$-vertex graph
can be solved in $1.1996^nn^{O(1)}$ time and polynomial space,
which even is faster than Robson's $1.2109^{n}n^{O(1)}$-time exponential-space algorithm published in 1986.
We also obtain improved algorithms for MIS in graphs with maximum degree 6 and 7,
which run in time of  $1.1893^nn^{O(1)}$ and $1.1970^nn^{O(1)}$, respectively.
Our algorithms are obtained by using fast algorithms for MIS in low-degree graphs
in a hierarchical way  and
making a careful analyses on the structure of bounded-degree graphs.

\vspace{5mm}\noindent {\bf Key words.}  \ \
Exact Algorithm, Independent Set, Graph, Polynomial-space, Branch-and-reduce, Measure-and-conquer, Amortized Analysis
\end{abstract}
\vspace{-0mm}

\section{Introduction}\label{sec_intr}

Over the last few decades, an extensive research has been done on exact exponential algorithms.
Many interesting methods and results have been obtained in this area, which can be found
in a nice survey by Woeginger~\cite{Woeginger:survey}
and a recent monograph by Fomin and Kratsch~\cite{Fomin:book}.
In the line of research on worst-case analysis of exact algorithms for NP-hard problems,
the \emph{maximum independent set} problem (MIS) is undoubtedly one of the most fundamental problems.
The problem is used to test the efficiency of some new techniques of exact algorithms and often
 introduced as the first problem in some textbooks and lecture notes of exact algorithms.
However, despite of a large number of contributions on exact algorithms and their worst-case analyses for MIS during the last 30 years, no published algorithm runs faster than the  $1.2109^nn^{O(1)}$-time exponential-space algorithm by Robson in 1986~\cite{Robson:IS}.
 Fomin and Kratsch say that `the running time of current branching algorithms for MIS
 with more and more detailed analyses seems to converge somewhere near $1.2^n$'~\cite{Fomin:book}.
 Researchers are interested in how fast we can exactly solve MIS and believe that some new techniques are
 required to get a further significant improvement.

\textbf{Related work.}
The first nontrivial exact algorithm for MIS is back to Tarjan and Trojanowski's $2^{n/3}n^{O(1)}$-time algorithm in 1977~\cite{Tarjan:IS}. Later, Jian obtained a $1.2346^{n}n^{O(1)}$-time algorithm~\cite{Jian:Is}.
Robson gave a
$1.2278^{n}n^{O(1)}$-time polynomial-space algorithm and a
$1.2109^{n}n^{O(1)}$-time exponential-space algorithm~\cite{Robson:IS}.
Robson also claimed better running times in a technical report~\cite{Robson:IS_1}.
A $1.0823^m n^{O(1)}$-time algorithm was introduced by Beigel in~\cite{Beigel:is}, where $m$ is the number of the edges in the graph.
Fomin \emph{et al.}~\cite{Fomin:is} introduced the ``measure-and-conquer" method and got a simple $1.2210^{n}n^{O(1)}$-time polynomial-space algorithm by using this method.
Also based on this method, Kneis \emph{et al}.~\cite{kneis:MIS} and Bourgeois \emph{et a}l.~\cite{Bourgeois:alg}
 improved the running time bound to $1.2132^nn^{O(1)}$ and $1.2114^nn^{O(1)}$ respectively,
which are the current fastest polynomial-space algorithms for MIS in published articles.
There is also a large amount
of contributions to MIS in degree-bounded graphs~\cite{Razgon:3IS,Furer:ISsparse,xiao:3ISwalcom,XN:3MIS,XN:4MIS-tr2,XN:5MIS}.
 Let MIS-$i$ mean  MIS in graphs with maximum degree $i$.
 Now MIS-3 can be solved in $1.0836^nn^{O(1)}$ time~\cite{XN:3MIS},
MIS-4 can be solved in $1.1376^nn^{O(1)}$ time~\cite{XN:4MIS-tr2},
MIS-5 can be solved in $1.1737^nn^{O(1)}$ time~\cite{XN:5MIS} and
MIS-6 can be solved in $1.2050^nn^{O(1)}$ time~\cite{Bourgeois:alg}, where all of them use only polynomial space.
The measure-and-conquer method is a 
 powerful tool to design or analyze exact algorithms. Most fast polynomial-space algorithms for MIS
are designed based on the method.
By combining this  method with a bottom-up method, Bourgeois \emph{et al}.~\cite{Bourgeois:alg} got the $1.2114^nn^{O(1)}$-time polynomial-space algorithm for MIS. Their algorithm is based on fast algorithms for MIS in low-degree graphs.

\textbf{Our contributions.}
In this paper, we will design a $1.1996^nn^{O(1)}$-time polynomial-space algorithm for MIS,
which is faster than Robson's $1.2109^{n}n^{O(1)}$-time exponential-space algorithm~\cite{Robson:IS} obtained in 1986.
We also show that MIS-6 and MIS-7 can be solved in $1.1893^nn^{O(1)}$ and $1.1970^nn^{O(1)}$ time, respectively.
Our algorithms use the measure-and-conquer method.
But the improvement is not  obtained  by studying more cases than previous algorithms.
Instead, we will introduce some new methods to reduce a large number of cases and make the algorithm and its analysis easy to follow.
Our algorithms also need to use our previous fast algorithms for MIS in low-degree graphs. The improvement is mainly obtained by using the following ideas:

\begin{enumerate}
\item
We exploit a divide-and-conquer method to get the improved algorithms for MIS in high-degree graphs based on fast algorithms for MIS in low-degree graphs.
In the method, we design an algorithm for MIS made of  two procedures.
One procedure is an algorithm to solve
MIS in graphs with maximum degree {\em at most} $i$.
The other procedure is to effectively deal with vertices of degree {\em at least} $i+1$
  in the graph.
We also use the idea to design fast algorithms for MIS in degree bound graphs.
Once an algorithm for MIS-$i$ is obtained, we design a procedure for eliminating
vertices of degree at least $i+1$
by reduction/branching operations,
which together with the algorithm for MIS-$i$ will give an algorithm for MIS-($i+1$).
Similar bottom-up ideas have been used in some previous algorithms,
 such as the algorithm for MIS in~\cite{Bourgeois:alg} and the algorithm for the parameterized vertex cover problem in~\cite{chen:VC2010}.
One advantage of our method is that, the divide-and-conquer method can combine the measure-and-conquer method well to design exact algorithms.
Then we can catch the properties of fast algorithms for MIS in low-degree graphs and propagates the improvement from instances of low-degree graphs to those of high-degree graphs.
\item
We devise a method that can reduce a huge number of case analyses in the algorithms
and then our algorithms become much easier to check the correctness.
This method is based on \refl{simplify_analysis} in Section~\ref{sec_highdegree}.
It can also be directly used  to reduce a large number of cases in the analysis of previous algorithms without     modifying the algorithms.
\item
We introduce a new branching rule, called ``branching on edges," to deal with edges between end-vertices
with many common neighbors,  for which the standard branching on a vertex of maximum degree has  not
 lead to a sufficiently high performance to improve the previous time bounds.
\end{enumerate}

\section{Preliminaries}\label{sec:preliminary}
\subsection{Notation system}
Let $G=(V,E)$ stand for a simple undirected graph with a set $V$ of vertices and a set $E$ of edges.
Let  $|G|$ denote $|V|$.
We will use  $n$ to denote $|V|=|G|$, $n_i$ to denote the number vertices of degree $i$ in $G$,
 and $\alpha(G)$ to denote the size of a maximum independent set of $G$.
The vertex set and edge set of a graph $G$ are denoted by $V(G)$ and $E(G)$, respectively.
 For simplicity,
 we may denote a singleton set $\{v\}$   by $v$.

For a vertex subset $X\subseteq V$
 in a graph $G$, we define the following notations.
Let $G-X$ denote the graph obtained from $G$ by removing $X$
together with edges incident on any vertex in $X$,
$G[X]=G-(V-X)$ be the  graph induced from $G$ by the vertices in $X$,
and $G/X$ denote the graph obtained from $G$ by contracting $X$
into a single vertex (removing self-loops and parallel edges).
Also we let $N(X)$ denote the set of all
vertices in $V-X$ that are adjacent to a vertex in $X$,
 and $N[X]=X\cup N(X)$.

For a vertex $v$ in a graph $G$ of   maximum degree $d$, we define the following notations.
Let
$\delta(v)=|N(v)|$ denote the degree of  $v$,
 $N_2(v)$ denote the set of vertices
with distance exactly $2$ from $v$, and $N_2[v]=N_2(v)\cup N[v]$.
Let $e_v$ denote the number of edges in the induced subgraph $G[N(v)]$ (i.e., $e_v=|E(G[N(v)])|$),
 let $f_v$ denote the number of edges between $N[v]$ and $N_2(v)$,
and let $q_v$  denote the number of vertices of degree $<d$ in $N_2(v)$.
Also define the {\em neighbor-degree}  $k_v$ of   $v$  to be the sequence $(k_1,k_2,\ldots,k_d)$
(where $d$ is the maximum degree of the graph)
of the number $k_i$ of degree-$i$ neighbors $u\in N(v)$.
Then  $\sum_{1\leq i\leq d} i k_i = \sum_{u\in N(v)}\delta(u) = \delta(v)+2e_v+f_v$.
We may denote $k_v=(k_3,k_4,k_5,k_6)$ when $k_1=k_2=0$ and $k_i=0$ for $i\geq 7$.

For each neighbor $u\in N(v)$ of $v$, we call
a vertex $z\in N(u)$ adjacent to $v$
(resp., not adjacent  to $v$) the {\em inner-neighbor} of $u$ at $v$
 (resp., {\em outer-neighbor} of $u$ at $v$).
Define the \emph{inner-degree}  (resp., {\em outer-degree}) of $u$ at $v$
to be the number of inner-neighbors (resp., outer-neighbors) of $u$ at $v$.
%
%

\subsection{Branching algorithms and the measure-and-conquer method}
Our algorithms use a branch-and-reduce paradigm.
We branch the current problem instance into several smaller instances to search a solution.
The iterative algorithm will create a search tree.
To scale the size of the instance, we need to select a measure for it.
A common measure of a graph problem is the number of vertices or edges in the graph.
By bounding the size of the search tree to a function of the measure,
we will get a running time bound related to the measure for the problem.
In MIS, a branching rule will branch on the current instance $G$ into
several instances $G_1$, $G_2$, $\ldots, G_l$ such that the measure
$\mu_i$ of each $G_i$ is less than the measure $\mu$ of $G$,
and a solution to $G$ can be
found in polynomial time if a solution to each of
the $l$ instances $G_1$, $G_2$, $\ldots, G_l$ is known.
Usually, $G_i$ ($i=1,2,\ldots, l$) are obtained by deleting some vertices in $G$.
We will use $C(\mu)$ to denote the worst-case size
of the search tree in the algorithm when the measure of the instance is at most $\mu$.
The above branch creates  the recurrence relation
 $C(\mu)\leq \sum_{i=1}^l C(\mu-\mu'_i)$, where $\mu'_i=\mu-\mu_i$.
 The largest root of the function $f(x)=1-\sum_{i=1}^l x^{-\mu'_i}$, denoted by $\tau(\mu'_1, \mu'_2, \ldots, \mu'_l)$,  is also called the \emph{branching factor} of the above recurrence relation.
 Let $\tau$ be the maximum branching factor among all branching factors in the search tree.
Then the size of the search tree is
 $C(\mu)=O(\tau^{\mu})$.
More details about the analysis and how to solve recurrences can be found in the monograph~\cite{Fomin:book}.

In some cases, the worst branch in the algorithm will not always happen. We can use the following idea of amortization to get better analysis.
Consider two branching operations $A$ and $B$
with recurrences $C(\mu)\!\leq\! C(\mu-t_{(A1)})+C(\mu-t_{(A2)})$ and $C(\mu)\!\leq\! C(\mu-t_{(B1)})+C(\mu-t_{(B2)})$
such that the branching operation $B$  leads to a better recurrence (with a smaller branching factor) than $A$ does, where the recurrence for the branching operation $A$ may be  the bottleneck
in the run time analysis of the algorithm.
Suppose that  branching operation  $B$ is always applied to the  subinstance $G_1$  generated by the first branch of $A$ in the algorithm.
In this case, we can obtain a better recurrence than that for $A$ if we derive a recurrence
by combining branching operation  $A$ and branching operation $B$ applied to $G_1$.
However, in general, there may be many branching operations $B_1,B_2,\ldots$ that can
be applied to $G_1$.
To ease such an analysis without generating all combined recurrences, we introduce
a notion of ``shift.''
To improve the branching factor of the recurrence for operation $A$,
 we transfer some amount from the measure decrease in the recurrence for operation $B$
to that for $A$ as follows.
We save an amount $\sigma >0$ of measure decrease from $B$
by evaluating the branch operation $B$
with  recurrence
\[C(\mu)\!\leq\! C(\mu-(t_{(B1)}-\!\sigma ))+C(\mu-(t_{(B2)}-\!\sigma )),\]
which is worser than its original recurrence.
The saved measure decrease $\sigma$ will be included into the recurrence for operation $A$ to
obtain
\[C(\mu)\!\leq\! C(\mu-(t_{(A1)}+\!\sigma ))+C(\mu- t_{(A2)} ).\]
The saved amount  is also called a \emph{shift}, where the best value for $\sigma$ will be
determined so that the maximum branching factor $\tau$ is minimized.
In our algorithm, we introduce one shift $\sigma$ in the analysis of our algorithm for MIS-6.

 To reduce the size of the search tree, we wish to find good branching rules, and
try to avoid  using bad branching rules with poor performance in designing algorithms.
 The selection of the measure is also an important issue in order to evaluate
how quickly problem instances can decrease after each branching operation.
 The measure-and-conquer method~\cite{Fomin:is} allows us to define a sophisticated way
of measuring the size of problem instances.
 In this method, we set a weight to each vertex in the graph according to the degree of the vertex
 (usually vertices of the same degree receive the same weight)
 and define the sum of the weights in the graph to be the measure.
 Note that when a vertex $v$ is deleted, we may decrease the measure not only from $v$
but also from the neighbors of $v$ since the degrees of the neighbors will decrease by $1$.
This yields an effect of amortizing branching factors from several different recurrences.
 Compared to the traditional measures, the weighted measure may catch more structural information of the
 graph and leads to a further improvement without modifying the algorithms;
 in fact, algorithms can be designed so that the target measure decreases as fast as possible
 before a final algorithm is proposed.
 Currently, the best exact algorithms for many NP-hard problems are designed by using this method.
 An important step in this method is to set  vertex weights as valuables.  We sometime solve a quasiconvex program
to determine the best values of them to minimize the maximum branching factor $\tau$.
 In this paper we also employ the branch-and-reduce paradigm as our algorithms  and
the measure-and-conquer method to analyze their run times.

\subsection{Reduction operations}
Before applying our branching rules, we may first apply some reduction rules to reduce some local structures, branching on which may lead to a bad performance.
Reduction rules can be applied in polynomial time to find a part of the solution or decrease the size of the instance directly.
Many nice reduction rules have been developed. In this paper, we only use three known reduction rules.

\bigskip \noindent
\textbf{Reduction by removing unconfined vertices}\\
A vertex $v$ in an instance $G$ is called {\em removable} if $\alpha(G)=\alpha(G-\! v)$.
A sufficient condition for a vertex to be removable has been studied
in  \cite{XN:3MIS}.
In this paper, we only use a simple case of the condition.
A neighbor $u\in N(v)$ of $v$ is called an {\em extending child} of $v$
if $u$  has exactly one outer-neighbor $s_u\in N_2(v)$ at $v$, where $s_u$ is also called an {\em extending grandchild} of $v$.
Let $N^*(v)$ denote the set of all extending children $u\in N(v)$ of $v$,
and $S_v$ be the set of all extending grandchildren $s_u$ ($u\in N^*(v)$) of $v$ together with $v$ itself.
We call $v$  {\em unconfined} if there is a neighbor $u\in N(v)$ which has no outer-neighbor or
 $S_v -  \{v\}$ is not an independent set
(i.e., some two vertices in $S_v\cap N_2(v)$ are adjacent)
\footnote{Unconfined vertices in  \cite{XN:3MIS} are defined in a more general way.}.
It is known in \cite{XN:3MIS} that any unconfined vertex is  removable.

\lem{unconfined}{{\rm \cite{XN:3MIS}}
For an unconfined vertex $v$  in graph $G$, it holds that
\vspace{-2mm}
 $$\alpha(G)=\alpha(G-\! v).$$ }

A vertex $u$ {\em dominates} another vertex $v$ if $N[u]\subseteq N[v]$,
where $v$ is called {\em dominated}.
We see that  dominated vertices  are  unconfined vertices.

\vspace{2mm}\noindent
\textbf{Reduction by folding complete $k$-independent sets}\\
We call a set $A=\{v_1,\ldots,v_k\}$ of $k$ degree-$(k+1)$ vertices
a {\em complete $k$-independent set} if they have
 common neighbors  $N(v_1)=\cdots =N(v_k)$.

\lem{complete-k-independent}{ {\rm \cite{XN:3MIS}}
For a complete $k$-independent set $A$, we have that
\vspace{-2mm}
\[ \alpha(G)=\alpha(G^\star )+k,\]
where
$G^\star=G/N[A]$ if $N(A)$ is  an independent set and
$G^\star=G- N[A]$ otherwise.
}

{\em Folding} a complete $k$-independent set $A$
 is to eliminate the set $N[A]$
from an instance in the above way.
In our algorithm, we only fold complete $k$-independent set with $k\leq 2$, since this operation is good enough for our analysis.
Folding a complete $1$-independent set $A=\{v\}$ consisting of a degree-2 vertex $v$ is also called {\em folding a degree-2 vertex $v$}.

\vspace{2mm}\noindent \textbf{Reduction by removing line graphs}\\
If a graph $H$ is the line graph of a graph $H'$,
then a maximum independent set of $H$ can be
obtained as the set of vertices that corresponds
the set of edges in a maximum matching in $H'$.
To reduce some worst cases, we need to remove the line graphs of 4-regular graphs, the line graphs of (4,5)-bipartite graphs
 (a bipartite graph with edges between two sets $V_1$ and $V_2$ is
a {\em $(d_1,d_2)$-bipartite graph} if every vertex in $V_i$ is of degree $d_i$ ($i=1,2$))
and the line graphs of 5-regular graphs.
A graph is the line graph of a 4-regular graph (resp., 5-regular graph) if and only if
the graph has only degree-$6$ vertices (resp., degree-8 vertices) and each of them is contained in two edge-disjoint cliques of size $4$ (resp., 5). A graph is the line graph of a (4,5)-bipartite graph if and only if
the graph has only degree-$7$ vertices and each of them is contained in two edge-disjoint cliques of size 4 and 5,  respectively.
More characterizations of line graphs can be found in~\cite{West:graphtheory}.
Removing line graphs of 4-regular  graphs (resp., (4,5)-bipartite graphs and 5-regular graphs)
is useful in the analysis of our algorithm for MIS-6 (resp., MIS-7  and MIS-8).

\dfn{reduced}{A graph is called a \emph{reduced graph}, if it contains none of
 unconfined vertices, complete $k$-independent sets with $k=1$ or $2$, and a component of a line graph of a 4-regular graph, a (4,5)-bipartite graph or a 5-regular graph.}

\begin{figure}[!h]
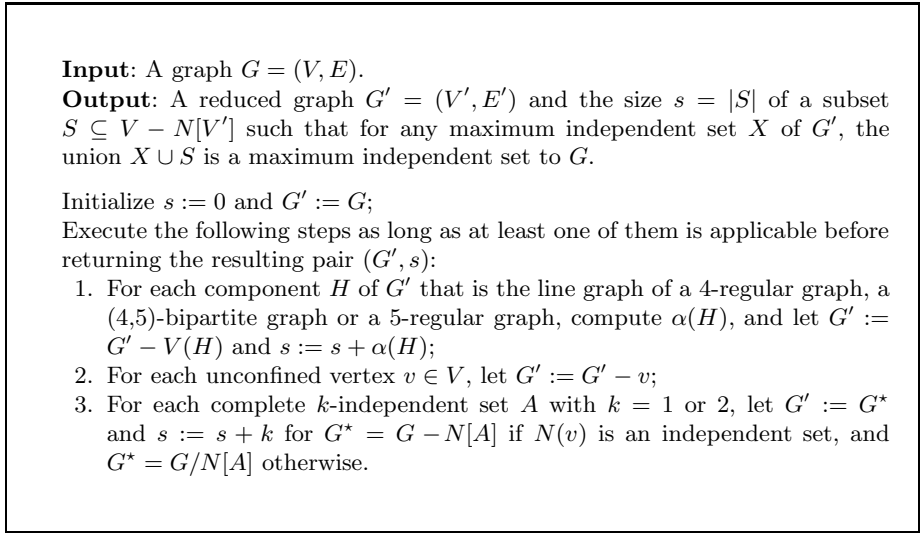

  \begin{center} \footnotesize
  \setbox4=\vbox{\hsize28pc
\noindent\strut
 \begin{quote} \vspace*{-0mm}
\textbf{Input}: A graph $G=(V,E)$. \\
\textbf{Output}: A reduced graph $G'=(V',E')$ and the size $s =|S |$ of a subset $S \subseteq V-N[V']$   such that
for any maximum independent set $X$ of $G'$, the union $X\cup S $ is a maximum independent set to $G$.\\

\vspace*{2mm}
Initialize $s:=0$ and $G':=G$; \\
Execute the following steps as long as at least one of them is  applicable before
returning the resulting pair $(G',s )$:
\begin{enumerate}
\item For each component $H$ of $G'$ that is the line graph of a 4-regular graph,  a (4,5)-bipartite graph or a 5-regular graph,
  compute $\alpha(H)$, and let $G' := G'-V(H)$ and $s := s +\alpha(H) $;
\item For each unconfined vertex $v\in V$,  let $ G' :=  G'-v $;
\item For each  complete $k$-independent set $A$ with $k=1$ or $2$, let
   $G' := G^\star$ and $s :=s +k$
for  $G^\star=G-\! N[A]$ if $N(v)$ is an independent set, and $G^\star=G/N[A]$ otherwise.
\end{enumerate}
\end{quote}
 \vspace*{-0mm} \strut}  $$\boxit{\box4}$$ \vspace*{-5mm}
\caption{Algorithm ${\tt reduce}(G,s)$} \label{alg_reduction}  \vspace{-0mm}
 \end{center}
\end{figure}

The algorithm in Figure~\ref{alg_reduction} is a collection of all above reduction operations.
When the graph is not a reduced graph, we can use the algorithm in Figure~\ref{alg_reduction} as a preprocessing to reduce it.
Notice that even if a graph of maximum degree $\theta$ is given as an instance
to MIS-$\theta$,
a vertex of degree  $d\geq \theta+1$ may be created by contraction of vertices
during an execution of  algorithm  ${\tt reduce}$.

 \vspace{-0mm}


\section{Divide-and-conquer method}\label{sec_dc}
We exploit a divide-and-conquer approach to design algorithms for solving MIS and MIS-$\theta$ $(\theta\geq 3)$.
In this method, we divide the class of instances of MIS or MIS-$\theta$ into two classes,
one consisting of instances of maximum degree at least  $j$ for some $3\leq j\leq \theta-1$,
and the other consisting of those of maximum degree at most  $j-1$.
For the first class of instances, we design a procedure that applies reduction/branching operations
until the maximum degree of the instance decreases to at most $j-1$.
Then we switch to an algorithm that solves the second class of instances, i.e., MIS-$(j-1)$.
We combine a procedure for the instances
of maximum degree at least $j$ with an algorithm for solving MIS-$(j-1)$
to obtain an algorithm for MIS or MIS-$\theta$.

However, sometimes it is not easy to analyze the  running time of the combined algorithms
since that a different measure may be used for the algorithm to each class.
We will introduce a method to effectively deal with this difficulty,
especially for the case where the measure is set as the sum of total weight of vertices in the graph.

We let $w_i\geq 0$
denote the weight of a degree-$i$ vertex in an instance $G$ of a class
and define the {\em measure} $\mu$  of the graph $G$ with $n_i$ degree-$i$ vertices $(i\geq 0)$ to be
\[ \mu(G) = \sum_{i} w_i n_i. \]
We may assign different values to $w_i$ 
under the conditions that $\mu(G)\leq n$ and any instance with $\mu(G)=0$ can be solved in polynomial time.
Hence if the measure never increases after any step of the algorithm and reduces after a branching operation, then
we can bound the size of search trees from above by a function $\tau^{\mu(G)}$ of $\mu(G)~(\leq n)$.


Let $A_i$ denote an algorithm that solves MIS-$i$  in a graph $G$ of maximum degree $\leq i$ in $(\tau_i)^{\mu_i(G)}|G|^{O(1)}$ time, where $\tau_i$ is a positive number and $\mu_i(G)=\sum_{1\leq j\leq i}w^{\langle i \rangle}_j n_j$ is the measure of $G$ (recall that $n_j$ is the number of degree-$j$ vertices in $G$ and $w^{\langle i \rangle}_j\geq0$ is the weight of a degree-$j$ vertex).
Let $B_{>i}$ denote a procedure  that branches on a graph $G$ of maximum degree $>i$ with branching factor $\tau'_{i}$ on measure $\mu_{i+1}(G)=\sum_{j\geq 1}w^{\langle i+1 \rangle}_j n_j$, where
$w^{\langle i+1 \rangle}_j\geq0$ is the weight of a degree-$j$ vertex in the procedure.
We have the following lemma for analyzing combined algorithms for MIS:

\lem{DCalg}{For an integer $i\geq 3$, let
$\lambda= \max\{ \frac{w^{\langle i \rangle}_j}{w^{\langle i+1 \rangle}_j}  \mid  0\leq j \leq i,~ w^{\langle i+1 \rangle}_j\neq 0 \}$ and ${\tau_{i+1}}=\max \{ \tau'_{i}, (\tau_i)^\lambda \}$.
Then MIS can be solved in $(\tau_{i+1})^{\mu_{i+1}(G)}|G|^{O(1)}$ time.}

\pf{We will construct an algorithm $A_{i+1}$ that solves MIS in $O^*((\tau_{i+1})^{\mu_{i+1}(G)})$ time. It iteratively applies the procedure $B_{>i}$ to branch when the graph has maximum degree $> i$, and calls the algorithm $A_i$ when the graph has maximum degree at most $i$. We analyze the running time of $A_{i+1}$.

In $A_{i+1}$, we use $\mu_{i+1}(G)$ as the measure (the same measure in $B_{>i}$). When the graph has a vertex of degree at least $i+1$, the algorithm can branch with branching factor $\tau'_{i}\leq\tau_{i+1}$.
When the graph becomes a graph of maximum degree at most $i$, the algorithm will execute $A_i$.
In this part, the algorithm uses time  $O^*((\tau_i)^{\mu_i(G_0)})$, where ${\mu_i(G_0)}=\sum_{1\leq j\leq i}w^{\langle i \rangle}_j n_j\leq \lambda\sum_{1\leq j\leq i}w^{\langle i+1 \rangle}_j n_j=\lambda \mu_{i+1}(G_0)$
(note that $n_j=0$ for $j>i$). This implies that the algorithm can always branch with branching factor $(\tau_i)^\lambda\leq \tau_{i+1}$ on measure $\mu_{i+1}(G)$ in this part.
Therefore, the algorithm $A_{i+1}$ runs in $O^*((\tau_{i+1})^{\mu_{i+1}(G)})$ time.
}\medskip


Here is an application of \refl{DCalg}.
In Sections~\ref{app_deg8} and \ref{sec_general}, we will show that MIS-8 can be solved in time $1.19951^{\mu_8(G)}|G|^{O(1)}$ time,
where $\mu_8(G)=0.65844n_3+0.78844n_4+0.88027n_5+0.95345n_6+0.98839n_7+n_8$,
 and that in a graph with maximum degree at least 9 we can branch with
branching factor $1.19749$ on the measure $\mu_9(G)=\sum_j n_j$.
In \refl{DCalg}, we have $\tau'_{8}=1.19749$, $\tau_8=1.19951$, and
$\lambda=\max \{0.65844, 0.78844, 0.88027, 0.95345, 0.98839, 1\}=1$.
Then MIS can be solved in $1.19951^nn^{O(1)}$ time.

In the above method, we let ${\tau_{i+1}}=\max \{ \tau'_{i}, (\tau_i)^\lambda \}$,
where $\tau'_{i}$ is decided by $B_{>i}$, $\tau_i$ is decided by $A_i$, and $\lambda$ is related
to the vertex weights in both of $B_{>i}$ and $A_i$.
So sometimes simple reductions on $\tau'_{i}$ or $\tau_i$ may not lead to improvement on the algorithm  $A_{i+1}$.
To get more properties and further improvements on the problem, in our algorithm,
we may not design  $A_i$ and $B_{>i}$ totally independently.
Instead, we will design $B_{>i}$ based on $A_i$ by considering the result (the values of $\tau_i$ and vertex weight)
of $A_i$ as some constraints to set the vertex weight in $B_{>i}$.

This divide-and-conquer method provides a way to solve MIS by solving two subproblems and
to design fast algorithms for MIS based on fast algorithms for MIS in low-degree graphs.
We will focus on the subalgorithm $B_{>i}$. Fast algorithms $A_i$ for MIS-$i$ with $i=3,4$ and $5$ can be found in references~\cite{XN:3MIS,XN:4MIS-tr2,XN:5MIS}.

In this paper, by using this divide-and-conquer method, first,  we design an algorithm for MIS-6
based on fast algorithm for MIS-5 in~\cite{XN:5MIS}, second, we design an
algorithm for MIS-7 based on the algorithm for MIS-6, third,
we design an algorithm for MIS-8 based on the algorithm for MIS-7, and
finally, we design an algorithm for MIS in general graphs based on the algorithm for MIS-8.
Our results are listed in  Table~\ref{table0}.

\begin{table}[htp]\centering
 \caption{Our algorithms designed by the divide-and-conquer method}
 \begin{center} \footnotesize
 \begin{tabular}{|l|l|l|l|}\hline
~Problem~ & ~Running time~ & Vertex weight & References \\ \hline \hline
~  MIS-6 &  ~$1.18922^nn^{O(1)}$ & ~ $(w_3,w_4,w_5 )= (0.49969, 0.76163, 0.92401)$
     & Section~\ref{sec:measure} \\ \hline
~   MIS-7 &  ~$1.19698^nn^{O(1)}$ & ~ $(w_3,w_4,w_5,w_6 )= $  & Section~\ref{app_deg7} \\
    &                  &  ~~~~~~~~(0.65077, 0.78229, 0.89060, 0.96384 )  &       \\ \hline
~   MIS-8  & ~$1.19951^nn^{O(1)}$  & ~ $(w_3,w_4,w_5,w_6,w_7 )=$ & Section~\ref{app_deg8}\\
~   (MIS)  &           &  ~~~~~~~~(0.65844, 0.78844, 0.88027, 0.95345, 0.98839) &   \\ \hline
 \end{tabular}
\end{center}
\label{table0}
\end{table}

\section{Branching on High-Degree Vertices}\label{sec_highdegree}
There is an easy way to deal with high-degree vertices.
We can simply branch on a high-degree vertex $v$ into two branches by including it to the solution set or not.
In the branch where $v$ is included to the solution,
$N[v]$ will be deleted from the graph since the neighbors of $v$ cannot be selected into the solution anymore.
If the degree of $v$ is higher, then the graph can be reduced more in this branch.
We extend the simple branch rule based on this following observation.
For a vertex $v$, there are only two possible cases:
(i) there is a maximum independent set of the graph which does not contain $v$; and
(ii) every maximum independent set of the graph contains $v$.
Recall here  the set $S_v$  of all extending grandchildren of $v$ together with $v$ itself.
As is shown in~\cite{XN:3MIS}, we see that for Case (ii),  $S_v$ is always contained
in any maximum independent set of the graph.
We get the following branching rule.

\emph{Branching on a vertex} $v$
 means generating two subinstances
by excluding $v$ from the independent set
or including  $S_v$  to the independent set.
In the first branch we will delete $v$  from the instance whereas
 in the second branch we will delete $N[S_v]$ from the instance.
\medskip

Branching on a vertex $v$ of maximum degree $d$ is one of the most fundamental operations in our algorithm.
We analyze this operation.
Throughout the paper, we use $\Delta w_i = w_i - w_{i-1}$ $(i\geq 3)$,
and  assume that
\eqn{new-min-delta}{
\mbox{$0\leq \Delta_{i+1}\leq \Delta_{i}$ ~~ $(i\geq 2)$; ~~
 $2\Delta_{\theta}\leq \Delta_{\theta-1}$ for MIS-$\theta$ $(\theta\in\{6,7,8\})$}
}
(these inequalites will be automatically satisfied with the optimized weights $w_i$ in our algorithms
to MIS-$\theta$).

%
Let $\Delta_{out}(v)$ and $\Delta_{in}(v)$ to denote the decrease of the measure of $\mu$
  in the branches of excluding $v$ and including $S_v$, respectively.
Then we get recurrence $C(\mu) = C(\mu \!-\!  \Delta_{out}(v) )+ C(\mu \!-\!  \Delta_{in}(v))$.
By letting $k_v=(k_3,k_4,\ldots,k_d)$ be the neighbor-degree of $v$,
we give more details about lower bounds on $\Delta_{out}(v)$ and $\Delta_{in}(v)$.

For the first branch, we get
\vspace{-2mm}
\[\Delta_{out}(v)= w_d+ \sum_{i=3}^d k_i \Delta w_i.  \]
Observe that, for a fixed neighbor-degree $k_v$, the decrease $\Delta_{out}(v)$ in the first branch is small when
 the neighbors of $v$ have higher degrees since $\Delta_i\geq \Delta_{i+1}$.

In the second branch, we will delete $N[S_v]$ from the graph.
Let $\Delta (\overline{N[v]})$ denote the decrease of weight of vertices in $V(G)- N[v]$
by removing $N[S_v]$ from $G$ together with possibly weight decrease attained by reduction operations applied to $G-\! N[S_v]$.
Then we have
 \[\Delta_{in}(v)\geq w_d+ \sum_{i=3}^d k_i w_i  +\Delta (\overline{N[v]}). \]
We observe that, for a fixed neighbor-degree $k_v$,  the decrease $\Delta_{in}(v)$ in the second branch
is determined by $\Delta (\overline{N[v]})$.

Then we can branch on a vertex $v$ of maximum degree $d$ with recurrence
\eqn{max-deg}{\begin{array}{*{20}l}
C(\mu)&=& C(\mu \!-\! \Delta_{out}(v))+ C(\mu \!-\! \Delta_{in}(v)) \\
&\leq &  C(\mu\!-\!(w_d+\sum_{i=3}^d k_i \Delta w_i))+ C(\mu\!-\!(w_d +\sum_{i=3}^d k_i w_i +\Delta (\overline{N[v]}))).
\end{array}}
A simple lower bound on $\Delta (\overline{N[v]})$ is obtained as follows.

\lem{degree-change}{For a vertex $v$ of maximum degree $d$, it holds
\[\Delta (\overline{N[v]})  \geq    (f_v \! +\! (f_v \! -\! |N_2(v)|) +q_v )\Delta w_d   \geq f_v\Delta w_d. \]}

\pf{
Let $\ell_z$ denote  the number of edges between $N(v)$ and $z$, where $f_v=\sum_{z\in N_2(v)} \ell_z$.
Since the degree of $z$ decreases by $\ell_z$ after removing $N[v]$ from $G$,
 we see that  $\Delta (\overline{N[v]}) \geq \sum_{z\in N_2(v)}(w_{\delta(z)}-w_{\delta(z)-\ell_z})$.
The number of degree-$d$ vertices in $N_2(v)$ is $|N_2(v)| -q_v$
(recall that $q_v$ is the number of vertices of degree $\leq d-1$ in  $N_2(v)$).
Since $\Delta_i\geq \Delta_{i+1}$ and $\Delta_{d-1}\geq 2\Delta_{d}$, we have
$\sum_{z\in N_2(v)}(w_{\delta(z)}-w_{\delta(z)-\ell_z})
= \sum_{z\in N_2(v):d(z)=d}(w_{d}-w_{d-\ell_z}) +  \sum_{z\in N_2(v):d(z)<d}(w_{d-1}-w_{d-1-\ell_z})
\geq  (|N_2(v)| -q_v) \Delta w_d +q_v\Delta w_{d-1} +  \Delta w_{d-1} \sum_{z\in N_2(v) } (\ell_z -1)
\geq  (|N_2(v)| -q_v) \Delta w_d +2\Delta w_{d}q_v + 2\Delta w_{d}(f_v \! -\! |N_2(v)|)
   =  (f_v \! +\! (f_v \! -\! |N_2(v)|) +q_v )\Delta w_d$, as required.
}\medskip

We here remark about some feature on the recurrence \refe{max-deg}.
In the recurrence \refe{max-deg},  usually
the measure decrease in the first branch of removing a vertex $v$ is much smaller than that in
the second branch, and the branching factor of the recurrence tends to be easily large
when the measure decrease in the first branch is small; i.e.,
 the neighbors of $v$ have higher degrees.
Another remark is a special effect of  the condition of $N^*(v)\neq \emptyset$
to the term $\Delta (\overline{N[v]})$.
Recall that  the second branch of including $S_v$ into the solution removes not only $N[v]$ but also
$N[S_v]-N[v]$.
This  provides a larger lower bound on  $\Delta (\overline{N[v]})$  than that in \refl{degree-change}
(see \refl{outside_decrease}(ii) for a detailed analysis).

As for branching on vertices of maximum degree in our algorithms,
we examine the recurrence \refe{max-deg} for all possible neighbor-degrees $(k_3, k_4,\ldots,k_d)$
to evaluate the branching factor precisely,
and show the existence of vertices that  attain a large value
  in  the lower bound  on $\Delta (\overline{N[v]})$ in \refl{degree-change}
based on a graph theoretical argument.

\medskip
Before closing this section, we propose a new method for knowing the maximum branching factor of recurrences
\refe{max-deg} over all neighbor-degrees of $v$.
Assume that
  we use a fixed lower bound on $\Delta (\overline{N[v]})$.
A straightforward method is to create a concrete recurrence for each neighbor-degree
$k_v=(k_3, k_4,\ldots,k_d)$ of $v$.
However, the number of neighbor-degrees $k_v=(k_3, k_4,\ldots,k_d)$ is  $(d+1)^{d-3}$
 (the number of integer solutions to the function $k_3+k_4+\cdots+k_d=d$).
Thus \refe{max-deg} actually consists of $(d+1)^{d-3}$ concrete recurrences.
We  introduce a technical lemma that can eliminate redundant recurrences to determine
the largest branching factor among a set of systematically generated recurrences.
With this, we can reduce
 the number of recurrences in \refe{max-deg}  from $(d+1)^{d-3}$ to only $d-2$.

\lem{simplify_analysis}{Let $C(x)=\tau^x$ for a positive $\tau>1$.
For  any nonnegative $p$, $\mu$,
$a_i$, $b_i$, $i=1,2,\ldots,\ell$ $(\ell\geq 1)$,
the maximum of
\[C(\mu- (\sum_{i=1,2,\ldots,\ell}k_i a_i \!+\!c))
+C(\mu- (\sum_{i=1,2,\ldots,\ell}k_i b_i\!+\!d))\]
over all $k_1,k_2,\ldots,k_{\ell}\geq 0$ subject to $k_1+k_2+\cdots+k_{\ell}=p$ is equal to
the maximum of
\[ C(\mu\!-\!(p a_i \!+\!c))+C(\mu - (pb_i\!+\!d)) \]
over all $i=1,2,\ldots,\ell$.
}

\pf{It suffices to show that for nonnegative $w$, $a_1$, $b_1$, $b_2$, $c$, $d\geq 0$,
it holds $C(\mu\!-\!(a_1\!+ a_2\!+\!c))+C(\mu\!-\!(b_1\!+ b_2\!+ d))\leq
\max\{C(\mu\!-\!(2a_1\!+ c))+C(\mu\!-\!(2b_1\!+ d)), C(\mu\!-\!(2a_2\!+ c))+C(\mu\!-\!(2b_2\!+ d))\}$.
The lemma can be obtained by applying this repeatedly.
Note that
 function $f(t)=\tau^{-(2a_1 (1-t)+2a_2 t+c-\mu)}+\tau^{-(2b_1 (1-t)+2b_2 t+d-\mu)}$
is convex since the second derivative  is nonnegative.
Hence  $f(0.5)\leq \max\{f(0),f(1)\}$ holds, as required.
}\medskip

By applying \refl{simplify_analysis}, in \refe{max-deg}, we only need to consider $d-2$ concrete recurrences with
neighbor-degrees
$(k_3, k_4,\ldots,k_d)=(d,0,\ldots,0), (0,d,0,\ldots,0), \cdots, $ $(0,\ldots,0,d)$,  respectively.
For example, if $d=6$, we can decrease the number of recurrences from  $7^4=2401$ to only $5$.
\refl{simplify_analysis} is introduced to simplify the analysis of recurrences for the first time. It can be
used to reduced thousands of recurrences in the analysis of previous algorithms for MIS, such as the algorithms in~\cite{kneis:MIS} and~\cite{Bourgeois:alg}.
Note that the authors of \cite{kneis:MIS} used a computer-added method to create all possible recurrences in the web
page~\cite{webpage}. There are more than 10 thousands recurrences listed.
By using \refl{simplify_analysis}, we need to generate a set of  about 50 recurrences, which is now easily checkable by hand.

\section{Branching on Edges}


As we have remarked in the previous section,
the maximum branching factor of recurrences \refe{max-deg}  becomes larger
when $N^*(v)=\emptyset$, $v$ has high degree neighbors,
and $\Delta (\overline{N[v]})$ is small.
By considering that $\Delta (\overline{N[v]})=f_v\Delta_{d}$ can hold in \refl{degree-change},
we wish to avoid branching on a vertex $v$ with a small $f_v$,
 particularly  when $N^*(v)=\emptyset$ and $k_d=d$.

Our solution to this situation is to introduce a new branching rule that can deal with the dense local graph $G[N(v)]$
caused by a small $f_v$ (a large $e_v$).
That is ``branching on edges.''

\vspace{2mm}\noindent \textbf{Branching on edges}
Two disjoint independent subsets $A$ and $B$ of vertices in a graph $G$
are called {\em alternative} if $|A|=|B|\geq 1$ and
there is a maximum independent set $S_G$ of $G$
which satisfies $S_G\cap (A\cup B)=A$ or $B$.
Let $G^\dagger$ be the graph obtained from  $G$ by removing
$A\cup B\cup (N(A)\cap N(B))$
and adding an edge $ab$ for  every two nonadjacent
vertices $a\in N(A)-N[B]$ and $b\in N(B)-N[A]$.

\lem{alternative}{{\rm \cite{XN:3MIS}} For alternative subsets $A$ and $B$ in a graph $G$,
  $\alpha(G)=\alpha(G^\dagger)+|A|$.
}

\lem{pair_branch}{
Let $v v'$ be an edge.
Then
\[\alpha(G)=\max\{\alpha(G-\{v,v'\}), \alpha(G^\dagger)+1\},\]
where
$G^\dagger$ be the graph obtained from  $G$ by removing
$\{v,v'\}\cup (N(v)\cap N(v'))$
and adding an edge $ab$ for  every two nonadjacent
neighbors $a\in N(v)-N[v']$ and $b\in N(v')-N[v]$.
 }

\pf{We easily observe that either
(i) every maximum independent set $S_G$ of $G$ satisfies $S_G\cap \{v,v'\}=\emptyset$; or
(ii) there is a  maximum independent set $S_G$ of $G$
such that $S_G\cap \{v,v'\}\neq\emptyset$.
In (i), we have $\alpha(G)= \alpha(G-\{v,v'\})$.
In (ii), sets $A=\{v\}$ and $B=\{v'\}$ are alternative in $G$, and we have $\alpha(G)= \alpha(G^\dagger)+1$
by \refl{alternative}.
}

\emph{Branching on an edge} $vv'$
 means generating two subinstances according to \refl{pair_branch}.
This is to either remove $\{v,v'\}$ from the graph $G$ or construct the graph $G^\dagger$ from $G-(\{v,v'\}\cup (N(v)\cap N(v')) )$
by making each pair $a\in N(v)-N[v']$ and $b\in N(v')-N[v]$ adjacent.
Branching on an edge may not always be very effective.
In our algorithms,
we will apply it to edges $vv'$ when $N(v)\cap N(v')$ is large, which are called ``short edges.''

We denote our algorithm for solving an instance of MIS-$\theta$ by  ${\tt mis}\theta(G)$.
The definitions of ``short edges'' in  algorithm ${\tt mis}\theta(G)$ are slightly different with $\theta$.
In  ${\tt mis}\theta(G)$, an edge $vv'$ in a reduced graph of maximum degree $\theta\in \{6,7,8\}$ is called {\em short} if \\
~(i)  $\delta(v)=6$, $\delta(v')\in \{5,6\}$ and $|N(v)\cap N(v')|\geq 3$ for $\theta=6$; and \\
~(ii)  $\delta(v)=\delta(v')=\theta$ and $|N(v)\cap N(v')| \geq 4$ for  $\theta\in \{7,8\}$.


A short edge is called {\em optimal} if $|N(v)\cap N(v')|-\delta(v')$ is maximized.
In our algorithms, we only branch on optimal
short edges in graphs of maximum degree 6, 7 and 8.

\section{Algorithms}
In this section, we describe our algorithms ${\tt mis}\theta(G)$, $\theta=6,7,8$, and then discuss MIS in general graphs.

\subsection{Algorithms for MIS-$6$, MIS-$7$ and MIS-$8$}

Our algorithm ${\tt mis}\theta(G)$, $\theta\in\{6,7,8\}$ is simple:
\begin{quote}
First keep branching on vertices of maximum degree $d>\theta$,
then keep branching on short edges, choosing an optimal one,
 and then  keep branching on vertices of maximum degree $\theta$,
choosing an ``optimal'' one.
During the execution, we switch to an algorithm for MIS-$(\theta-1)$
whenever  the maximum degree of the graph becomes smaller than $\theta$.
\end{quote}
See Figure~\ref{misgeneral} for their descriptions.
In the rest of this section, we describe which vertices of maximum degree should be
chosen as ``optimal''   vertices.
When no short edge is left in a graph $G$  of maximum degree $\theta\in \{6,7,8\}$,
the inner-degree of each neighbor $u$ of a degree $\theta$-vertex
 $v$ is at most 2 for $\theta=6$
       and 3 for $\theta\in \{7,8\}$
(otherwise $vu$ would be a short edge).
In particular, for every degree $\theta$-vertex $v$ with  $N^*(v)=\emptyset$,
the outer-degree of every neighbor $u\in N(v)$
is at least 2 at $v$, and  we have
 $f_v\geq 2\delta(v)+k_{\delta(v)}$ for $\theta\in\{6,7\}$;
 $f_v\geq 2\delta(v)+2k_{\delta(v)}$ for $\theta=8$.
As we have remarked,
we know that the branching factor of recurrence \refe{max-deg}  tends to be larger
 when $N^*(v)$ is  empty and the neighbor-degree $k_v$ consists of higher degrees.
As a vertex $v$ to branch on first should be one with $N^*(v)\neq \emptyset$,
or with a neighbor-degree $k_v$ which is lexicographically small.
When $k_v$ is close to $(k_3=0,0,\ldots,0,k_\theta=\theta)$, we choose a vertex
that attains a large value in  the lower bound
$(f_v \! +\! (f_v \! -\! |N_2(v)|) +q_v )$ in  \refl{degree-change}.
This is our priority for selecting vertices of maximum degree $\theta$.
We define ``optimal'' vertices for each ${\tt mis}\theta(G)$, $\theta=6,7,8$ according to this.

In a reduced graph of maximum degree 6 in MIS-6,  a degree-6 vertex $v$ is called {\em optimal}
if at least one of the following (i)-(vi)  is holds:\\
(i) $k_3\geq 1$ or  $k_6\leq 3$; \\
(ii) $k_6=4$ and $k_5\leq 1$; \\
(iii) $k_6=4$, $k_5=2$ and $f_v+(f_v \! -\! |N_2(v)|)+q_v \geq 17$; \\
(iv)   $k_6=5$, $k_4=1$ and $f_v+(f_v \! -\! |N_2(v)|)+q_v \geq 18$; \\
(v) $k_6=5$, $k_5=1$ and $f_v+(f_v \! -\! |N_2(v)|)+q_v \geq 19$; and \\
(vi)  $k_6=6$ and $f_v+(f_v \! -\! |N_2(v)|)+q_v \geq 22$.

\smallskip
In a reduced graph of maximum degree 7 in MIS-7,
a degree-7 vertex $v$ is called {\em optimal}
if at least one of the following (i)-(iv) is holds:\\
(i) $N^*(v)\neq \emptyset$;\\
(ii) the vertex $v$ has at most $5$ degree-$7$ neighbors ($k_{7}\leq 5$);\\
(iii) $k_7=6$ and $f_v+(f_v \! -\! |N_2(v)|)\geq 22-2k_3-k_4$; and\\
(iv) $k_7=7$  and $f_v+(f_v \! -\! |N_2(v)|)\geq 26$.

\smallskip
In a reduced graph of maximum degree 8 in MIS-8,
a degree-8 vertex $v$ is called {\em optimal}
if (i) $k_8\leq 7$ or (ii) $k_8=8$ and $f_v+(f_v \! -\! |N_2(v)|)\geq 36$.

Note that in the definitions of optimal vertices
in graphs of maximum degree 7 and 8, we do not need to use $q_v$.

\lem{general-optimal}{Let $G$ be a reduced graph of maximum degree $\theta$ in MIS-$\theta$ $(\theta\in\{6, 7,8\}).$
If $G$ has no short edges, then $G$ has at least one optimal vertex.}

In order to focus on the mechanism of our algorithms first, we  move the proof of this purely
analytical lemma to Section~\ref{sec_proof}.

 \vspace{-0mm}\begin{figure}[!h]
  \begin{center} \footnotesize
 \setbox4=\vbox{\hsize28pc
\noindent\strut
\begin{quote}
\textbf{Input}: A graph $G$.  \\
\textbf{Output}: The size of a maximum independent set in $G$.\\

\vspace*{2mm}
\begin{enumerate}
\item Reduce the graph by $(G,s):={\tt reduce}(G,0)$,
 and let $d$ be the maximum degree of $G$.
\item \textbf{If} \{$d\geq (\theta+1)$ \},
pick up a vertex $v$ of maximum degree $d$, and \textbf{return}
$s+\max\{{\tt mis}\theta(G-\! v), |S_v|+{\tt mis}\theta(G-\! N[S_v])\}$.
\item \textbf{Elseif}\{$d=\theta$ and $G$ has a short edge\},
 pick up an optimal short edge $vv'$, and
  \textbf{return} $s+\max\{{\tt mis}\theta(G-\! \{v,v'\}), 1+{\tt mis}\theta(G^\dagger) \}$.
\item \textbf{Elseif} \{$d=\theta$ ($G$ has no short edges)\},
 pick up an optimal degree-$\theta$ vertex $v$, and
 \textbf{return} $s+\max\{{\tt mis}\theta(G-\! v), |S_v|+{\tt mis}\theta(G-\! N[S_v])\}$.
\item \textbf{Else} \{The maximum degree of $G$ is smaller than $\theta$\},
 use our algorithm for MIS-$(\theta\! -\! 1)$ 
to solve the instance $G$ and \textbf{return} $s+\alpha(G)$,
where the algorithm for MIS-5 is in~\cite{XN:5MIS}.
\end{enumerate}

\vspace{2mm}
\textbf{Note}: With a few modifications, the algorithm
can deliver a maximum independent set.
\end{quote}  \vspace*{-0mm} \strut}  $$\boxit{\box4}$$ \vspace*{-5mm}
\caption{Algorithms ${\tt mis}\theta(G)$} \label{misgeneral} 
 \end{center}
\end{figure}

In our algorithms ${\tt mis}\theta$ $(\theta=6,7,8)$,
we set the vertex weight $w_i$ $(i\geq 3)$ as follows (recall that $w_1=w_2=0$):
For $3\leq i\leq \theta-1$,   $w_i$  is set as that in Table~\ref{table0}; and
\eqn{high-degree-weight}{
 w_{\theta}=1;~~
 w_{i}=w_{\theta} + (i-\theta)\Delta w_{\theta} ~~~ i\geq \theta+1. }   
To simplify our analyses, the weights of vertices of degree $\geq \theta+1$ are allowed to be greater than 1.
Recall that a vertex of degree  $\geq \theta+1$ may be created after applying reduction rules.
As will be observed, we create vertices of degree  $\geq \theta+1$ only when the measure does not increase.
This ensures that  the running time bound of our algorithms still can be expressed by
$\tau^n n^{O(1)}$ with the largest branching factor $\tau$.

 \lem{branch-general}{With the above vertex weight setting, each recurrence generated
 by the algorithm ${\tt mis}6(G)$ $($resp., ${\tt mis}7(G)$ and ${\tt mis}8(G))$
 in Figure~\ref{misgeneral} has a branching factor not greater than $1.18922$ $($resp., $1.19698$ and $1.19951)$.}

We will give  detailed analysis of our algorithms ${\tt mis}6(G)$,  ${\tt mis}7(G)$ and ${\tt mis}8(G)$
 in Sections~\ref{sec:measure}, \ref{app_deg7} and \ref{app_deg8}, respectively,
 and then complete a proof of \refl{branch-general}.
Since the measure $\mu$ is not greater
than the number $n$ of vertices in the initial graph
 in ${\tt mis}6(G)$, ${\tt mis}7(G)$ and ${\tt mis}8(G)$, we establish the next.

\thm{result678}{A maximum independent set in an $n$-vertex graph with maximum degree $\theta\in \{6,7,8\}$
   can be found in time of $1.1893^nn^{O(1)}$ for $\theta=6$,
 $1.1970^nn^{O(1)}$ for  $\theta=7$, and  $1.1996^nn^{O(1)}$ for $\theta=8$, respectively.}

\subsection{MIS in general graphs}\label{sec_general}
Our algorithm for MIS in general graphs is also simple.
It only contains two steps: Keep branching on a vertex of maximum degree while the degree of the graph is 9 or lager,
and invoke our algorithm ${\tt mis}8$ for MIS-8 whenever the maximum degree of the graph becomes  less than 9.
For the procedure for dealing with vertices of degree $\geq 9$,
we set the measure as the number of vertices in the graph (the weight of each vertex is 1).
Then we can get the following recurrence:
\eqn{deg>9}{
C(\mu) &\leq &  C(\mu-1) + C(\mu-10),
}
which has a branching factor $1.19749$, better than 1.19951 for MIS-8.
The analysis in Section~\ref{sec_dc} shows that MIS in general graphs can also be solved in $1.19951^nn^{O(1)}$ time.

\medskip
\thm{resultall}{A maximum independent set in an $n$-vertex graph can be found in $1.1996^nn^{O(1)}$ time and polynomial space.}

\section{Analysis of ${\tt mis}6(G)$}\label{sec:measure}

For MIS-6, we first give some properties of the subgraph $G[N(v)]$ of the neighbors of an optimal vertex $v$,
and then derive recurrences for all branching operations in ${\tt mis}6(G)$.

\subsection{Weight setting}

Recall that, for MIS-6,
we assume that $w_0=w_1=w_2=0\leq w_3\leq w_4\leq w_5\leq w_6= 1\leq w_7\leq w_8\leq \cdots , $
and the values of $w_3$, $w_4$ and $w_5$ will be determined
after we analyze how the measure changes after each step of the algorithm.

To simplify our analysis, we assume that
\eqn{new-1}
{6\Delta w_6\leq w_3,}
where this condition is satisfied by the optimized values in  Table~\ref{table0}.
\invis{
 c( )=6*(x(6)-x(5))-x(3); 
 }
%
\invis{
 c( )= 2*(x(6)-x(5)) -(x(5)-x(4)); 
 c( )= (x(5)-x(4)) -(x(4)-x(3)) ; 
 c( )= (x(4)-x(3)) -x(3). 
}

We impose the next constraint in order to ensure that contracting vertices will never
increase the measure:
\eqn{weight-merge}{w_i+w_j\geq w_{i+j-2},~~~ 3\leq i, j\leq 6.}
\invis{
 c( )=x(4) -x(3)-x(3) ; 
 c( )= x(5)-x(3)-x(4)   ; 
 c( )= x(6)-x(4)-x(4) ; 
 c( )= x(6)-x(5)-x(3)   ; 
 c( )= x(6)+(x(6)-x(5))-x(5)-x(4)  ; 
 c( )= x(6)+2*(x(6)-x(5))-x(5)-x(5)  ; 
}%
\lem{weight-sum}{$w_i+w_j\geq w_{i+j-2}$ holds for all $i,j\geq 1$.
}

\pf{If $i$ or $j$ is at most $2$, say $i\leq 2$,  then $w_i+w_j =w_j\geq w_{i+j-2}$.
Let $i,j\geq 3$.
For $3\leq i,j \leq 5$, we have $w_i+w_j\geq w_{i+j-2}$
by \refe{weight-merge}.
Let  at least one of $i$ and $j$, say $i$, is at least 6.
Then
we have that $i+j-2\geq 7$ and
 $w_{i+j-2}=w_i+(j-2)\Delta w_6$
by the definition of $w_k$ $(k\geq 7)$.
Since $w_j\geq (j-2)\Delta w_6$, this
  implies that $w_{i+j-2}=w_i+(j-2)\Delta w_6\leq w_i+w_j$.
}

\lem{reduction_123}{
The measure $\mu$ of a graph $G$ never increases
 in RG$(G,s)$ $($Step 1 of $\mathrm{mis}6(G))$. Moreover, in a graph of maximum degree $d$ and minimum degree $\geq 3$, the measure $\mu$ decreases by at least $2\Delta w_d$
after applying RG$(G,s)$ if the maximum degree of the graph decreases by at least one.
}

\pf{RG$(G,s)$ contains  three reduction steps. In Step~1, when a component $H$ of the line graph of a 4-regular graph,  a (4,5)-bipartite graph or a 5-regular graph is removed,
the measure never increases. If a vertex of maximum degree is removed in this step, then the measure decreases by at least $w_d\geq 2\Delta w_d$.
In Step~2, when an unconfined vertex is removed, the measure will not increase, since the vertex weight is monotonic with the degree of the vertices. If a vertex of maximum degree is
reduced, then  the measure decreases either by $w_d\geq 2\Delta w_d$ from this vertex or by  $2\Delta w_d$ from this vertex and a neighbor of it.
In Step~3,  folding a complete $k$-independent set ($k=1$ or $2$) is applied.
By \refl{weight-sum}, we know that the measure will never increase in this step.
Note that in this step,
either  $N[A]$ is removed or a graph $G/N[A]$ is generated by contracting $N[A]$.
For the former case, the measure decreases by at least $w_3\geq 2\Delta w_d$.
For the latter case,
we can see that either the measure decreases by at least $2\Delta w_d$ directly or the new created vertex has a degree not less than the degree of any vertex in $N[A]$ and all vertices other than $N[A]$ keep the same degree, which means that the maximum degree of the graph does not decrease. This completes the proof.
}


\medskip
Next  we will analyze the recurrence of each branching step of ${\tt mis}6(G)$.

\subsection{Branching on vertices of maximum degree in Step~2 of ${\tt mis}6(G)$}\label{sec:max-degree}

We will derive recurrences for branchings in Step~2 of ${\tt mis}6(G)$.
Let $G$ be a reduced graph after Step~1 of ${\tt mis}6(G)$.
The next property holds for every vertex $v$ in $G$.

\lem{outside_decrease}{Let $v$ be a vertex in a reduced graph $G$, and $f_v$ denote
the number of edges between $N(v)$ and $N_2(v)$, where  $f_v\geq  \delta(v)$.\\
{\rm (i)}
If $N^*(v)=\emptyset$, then
 $f_v\geq 2\delta(v)$ and  $\Delta(\overline{N[v]})\geq 2\delta(v)\Delta w_6$. \\
{\rm (ii)}
If $N^*(v)\neq \emptyset$,
   then $\Delta(\overline{N[v]}) \geq
  \min \{2w_3 , w_3+2(\delta(v)-3)\Delta w_6\}$.
}

\pf{In general, we have  $\Delta(\overline{N[v]})\geq f_v\Delta w_6$ since  $\Delta w_i\geq \Delta w_6$
 by \refe{new-min-delta} and
$6\Delta w_6\leq w_3 $ by \refe{new-1}.
Each neighbor of $v$ has a neighbor in $N_2(v)$ and $f_v\geq  \delta(v)$,
since otherwise it would dominate some other neighbor of $v$.

\noindent
(i) If $N^*(v)=\emptyset$, then
 there are at least $2\delta(v)$ edges between $N(v)$ and $N_2(v)$ and then $f_v\geq  2\delta(v)$.

\noindent
(ii) Assume that $N^*(v)\neq \emptyset$; i.e., $S_v-\{v\}\neq\emptyset$.
If $|S_v-\{v\}|\geq 2$, then clearly $\Delta(\overline{N[v]})\geq \sum_{t\in (S_v-\{v\})}w_{\delta(t)}\geq 2w_3$.
Assume that $|S_v-\{v\}| =1$ and $u$ is the unique vertex in $S_v-\{v\}$. Each vertex in $N^*(v)$ is adjacent to $u$ and
each vertex in $N(v)-N^*(v)$ is adjacent to at least two vertices in $N_2(v)$. Let $d^*=|N^*(v)|$. Then
$\Delta(\overline{N[v]}) \geq  w_{\delta(u)}+2(\delta(v)-d^*)\Delta w_6$. Note that $\delta(u)\geq \max\{3, d^*\}$.
We know that $\Delta(\overline{N[v]}) \geq  \min\{ w_{i}+2(\delta(v)-i)\Delta w_6\mid 3\leq i\leq \delta(v)\}
\geq w_3+2(\delta(v)-3)\Delta w_6$ (by \refe{new-min-delta}).
}\medskip

In particular, for vertices $v$ with $\delta(v)\geq 7$, we obtain
 $\Delta (\overline{N[v]})\geq
  \min \{2\delta(v) \Delta w_6,$ $ 2w_3 , w_3+ 2(\delta(v)-3)\}\geq   12\Delta w_6$
  by \refe{new-1} and \refe{new-min-delta}.

Now we derive recurrence of branching on a vertex of degree $\geq 7$ in Step~2 of ${\tt mis}6(G)$.
To evaluate the largest branching factor  of the recurrences \refe{max-deg} with the lower bound $\Delta (\overline{N[v]})$
for all $d\geq 7$,
we only need to consider those for vertices with no neighbors of degree $d\geq 7$,
 since $\Delta w_6=\min\{\Delta w_i\}$ for all $i\geq 3$ and
$w_{i+1}\geq w_i$ for all $i\geq 3$.
Furthermore, this means that we only have to consider the case of $d=7$
 in cases of $k_i=7$ $(3\leq i \leq 7)$
by \refl{simplify_analysis}.
Thus we get  recurrences:
\[
 \begin{array}{*{20}l}
C(\mu) &\leq &  C(\mu\!-\!(w_d+\sum_{i=3}^d k_i \Delta w_i))
+ C(\mu\!-\!(w_d +\sum_{i=3}^d k_i w_i + \Delta (\overline{N[v]}))) \\
&\leq &   \max_{3\leq i\leq 7} [C(\mu\!-\!(w_7\!+\!7\Delta w_i)) + C(\mu\!-\!(w_7 \!+\!7 w_i\!+\!  12\Delta w_6  ))].
\end{array}
\]

These recurrences will not leads to the largest branching factor.
In fact, two of the recurrences for branching on short edges in Step~3 of  algorithm ${\tt mis}6(G)$
 (\refe{deg6:K23:4 i j} with $i=6$ and $j=6$
and \refe{deg6-6:K23:i>3} where $i=6$ and $j=6$) will be the worst recurrences.
Since we know that a vertex of degree $\geq 7$ always will be created after the second branch
in such short edge branching,
we here save a shift $\sigma>0$ from the recurrence for branching on  vertices of degree $\geq 7$
so that the shift $\sigma>0$ will be included into the recurrences for the short edge branchings.
Then in this step we use the following recurrences indeed:
\eqn{deg>=7:K24:i}{
C(\mu) &\leq & \max_{3\leq i\leq 7} [C(\mu\!-\!(w_7\!+\!7\Delta w_i-\sigma)) +
C(\mu\!-\!(w_7 \!+\!7 w_i\!+\!  12\Delta w_6 -\sigma ))].
}
\invis{
 c( )=-1+x(1)^(-(x(7)+7*(x(7)-x(6))-x(10) )) +x(1)^(-( x(7)+7*x(7)  +12*(x(6)-x(5))-x(10) )) ;
 c( )=-1+x(1)^(-(x(7)+7*(x(6)-x(5))-x(10) )) +x(1)^(-( x(7)+7*x(6)  +12*(x(6)-x(5))-x(10) )) ;
 c( )=-1+x(1)^(-(x(7)+7*(x(5)-x(4))-x(10) )) +x(1)^(-( x(7)+7*x(5)  +12*(x(6)-x(5))-x(10) )) ;
 c( )=-1+x(1)^(-(x(7)+7*(x(4)-x(3))-x(10) )) +x(1)^(-( x(7)+7*x(4)  +12*(x(6)-x(5))-x(10) )) ;
 c( )=-1+x(1)^(-(x(7)+7*(x(3)     )-x(10) )) +x(1)^(-( x(7)+7*x(3)  +12*(x(6)-x(5))-x(10) )) ;
}%

\subsection{Branching on short edges in Step~3 of ${\tt mis}6(G)$} \label{sec:bipartite}

We derive recurrences for branching on optimal short edges  $vv'$ in Step~3 of ${\tt mis}6(G)$.
Let $v$ be a degree-6 vertex,  $d'=\delta(v')\in \{5,6\}$, and
$k\in \{3,4\}$ be the number of   common neighbors of $v$ and $v'$.
Denote    $N(v)-\{v'\}=\{u_i\mid i=1,2,3,4,5\}$,
 $N(v')-\{v\}=\{u'_i\mid i=1,2,\ldots,d'-1\}$, where $u_i=u'_i$,
 $1\!\leq\! i\!\leq\! k$ are the common neighbors,
and let $i^*$ denote the number of degree-3 vertices $u\in \{u_1,\ldots,u_k\}$,
where we assume that for each $i\leq i^*$, $u_i$ is a degree-3 neighbor of $v$.
Let $X=\{v,v'\}\cup (N(v)\cap N(v'))$.
We distinguish three cases: (i)  $d'=6$ and $k=4$; (ii)  $d'=5$ and $k=3$; and (iii) $d'=6$ and $k=3$.
By the optimality of the selected short edge $vv'$ in this step, we know that: when $vv'$ satisfies (ii) then no short edge satisfies (i);
and when  $vv'$ satisfies (iii) then no short edge satisfies (i) or (ii).
This is the reason why we need to define optimal short edges.

\medskip
\noindent
Case (i)  $d'=6$ and $k=4$:
The first branch of deleting vertices $v$ and $v'$ decreases
the degree of $u_i$ ($1\!\leq\! i\!\leq\! 4$) by two and
that of $u_5$ and $u'_5$   by one.
Each degree-3 neighbor $u_i$ $(i\leq i^*)$ will be a degree-1 vertex in $G-\{v,v'\}$
and its unique neighbor $z_i\in N_2(v)\cap N_2(v')$ will be removed since it is an unconfined vertex,
where $z_i\not\in \{u_5,u'_5\}$ (otherwise $u_i$ would dominate $z_i~(=u_5,u'_5)$)
and
 $z_\ell\neq z_{\ell'}$ for $1\leq \ell<\ell'\leq i^*$
 (otherwise $(u_\ell,u_{\ell'})$ would a complete $k$-independent set which must have been reduced in ${\tt reduce}$).
Hence in the first branch  the measure decreases by at least
$2w_6+\sum_{1 \leq\! i \leq\! 4}(w_{\delta(u_i)}-w_{\delta(u_i)-2})
+\Delta w_{\delta(u_5)} +\Delta w_{\delta(u'_5)}+i^*w_3$, where $i^*w_3$ is from deleting $z_i$.

In the second branch we delete $X=\{v,v',u_1,u_2,u_3,u_4\}$  to construct  graph $G^\dagger$,
joining  $u_5$ and $u'_5$ with a new edge if they are not adjacent in $G$.
Note that $G$ has at least four edges between $X$ and $V-X$  other than edges $vu_5$ and $v'u'_5$.
The second branch
decreases the weight of vertices in $V-(X\cup\{u_5,u'_5\})$ by at least $4\Delta w_6$
 even after an edge $u_5 u'_5$ is introduced
(since edges $vu_5$ and $v'u'_5$ are not included to evaluate the measure decrease).
Hence in the second branch,  the measure
decreases by at least
$2w_6+\sum_{1 \leq\! i \leq\! 4} w_{\delta(u_i)} + 4\Delta w_6$.
By \refl{simplify_analysis}, we only need to consider the following four recurrences
each of which corresponds to the case of $\delta(u_1)=\delta(u_2)=\delta(u_3)\in \{3,4,5,6\}$:
%
\eqn{deg6:K24:i}{
\begin{array}{*{20}l}
   C(\mu) &  \leq  & C(\mu\!-\!(2w_6+4(w_i\!-\!w_{i-2}) +2\Delta w_6))\\
&&    + C(\mu\!-\!(2w_6+4w_i+4\Delta w_6 )) ~~(i=4,5,6);
\end{array}}
and
\eqn{deg6:K24:3}{
   C(\mu)  \leq   C(\mu\!-\!(2w_6+8w_3 +2\Delta w_6))  + C(\mu\!-\!(2w_6+4w_3+4\Delta w_6 )). }
\invis{
%
c( )= -1+x(1)^(-(2*x(6)+4*(x(6)-x(4))+2*(x(6)-x(5)) )) + x(1)^(-(2*x(6)+4*x(6)+4*(x(6)-x(5)) ));
c( )= -1+x(1)^(-(2*x(6)+4*(x(5)-x(3))+2*(x(6)-x(5)) )) + x(1)^(-(2*x(6)+4*x(5)+4*(x(6)-x(5)) ));
c( )= -1+x(1)^(-(2*x(6)+4*x(4)       +2*(x(6)-x(5)) )) + x(1)^(-(2*x(6)+4*x(4)+4*(x(6)-x(5)) ));
c( )= -1+x(1)^(-(2*x(6)+8*x(3)       +2*(x(6)-x(5)) )) + x(1)^(-(2*x(6)+4*x(3)+4*(x(6)-x(5)) ));
}

Next, we assume that there is no short edges $vv'$ with $\delta(v)=\delta(v')=6$ and $|N(v)\cap N(v')|=4$.
Then  the outer-degree of every degree-6 neighbor of a degree-6 vertex is at least 2.
\medskip

\noindent
Case (ii)   $d'=5$ and $k=3$:
Analogously with Case (i), the first branch decreases the measure by at least
$w_6+w_5+\sum_{i=1,2,3}(w_{\delta(u_i)}-w_{\delta(u_i)-2})
+\Delta w_{\delta(u_4)}+\Delta w_{\delta(u_5)} +\Delta w_{\delta(u'_4)} +i^*w_3
\geq w_6+w_5+\sum_{i=1,2,3}(w_{\delta(u_i)}-w_{\delta(u_i)-2})
+2\Delta w_6 +\Delta w_{j}+i^*w_3$, where $j=\delta(u'_4)\in \{3,4,5,6\}$.

We consider the second branch of deleting  $X=\{v,v',u_1,u_2,u_3\}$ from $G$
to construct $G^\dagger$ by adding edges $u_4u'_4$ and $u_5u'_4$, if necessary.
Let $p$ be the number of degree-6 vertices in $\{u_1,u_2,u_3\}$, where each degree-6 neighbor of $v$ has
outer-degree at least 2.
Let $L$ denote the number of
edges in $G$ between $X$ and $V-X$  other than the three edges $vu_4$, $vu_5$ and $v'u'_4$,
where $L\geq 3+p$.
Recall that $j$ is the degree of $u'_4$ in $G$.
Then the degree of $u'_4$ in $G-X$ is $j-\ell-1$,
where  $\ell$ is the number of edges between   $u'_4$ and $\{u_1,u_2,u_3\}$.
Then the degree of $u'_4$ in $G^\dagger$ will be at most $j-\ell+1$,
and the weight change at vertex $u'_4$ from $G$ to $G^\dagger$ is at least
\[w_j\!-\! w_{j-\ell+1} = - (w_{j+1}\!-\! w_j) +(w_{j+1}\!-\! w_{j-\ell+1})
~(\geq  - (w_{j+1}\!-\! w_j)  + \ell \Delta w_6).\]
Recall that $L\geq \ell$.
Hence the decrease of the measure in the second branch is at least
$w_6+w_5+\sum_{i=1,2,3} w_{\delta(u_i)} + L\Delta w_6 - (w_{j+1}\!-\!w_j)
\geq  w_6+w_5+\sum_{i=1,2,3} w_{\delta(u_i)} + (3+p)\Delta w_6 - (w_{j+1}\!-\!w_j)$.
By \refl{simplify_analysis}, we only need to consider the following  recurrences:
\eqn{deg6:K23:4 i j}{
\begin{array}{*{20}l}
   C(\mu) &  \leq  & C(\mu\!-\!(w_6+w_5+3(w_i\!-\!w_{i-2})+2\Delta w_6+\Delta w_j))\\
&&    + C(\mu\!-\!(w_6+w_5+3w_i+ (3+p)\Delta w_6 \!-\!(w_{j+1}\!-\!w_j) )),

\end{array}}
where $4\leq i\leq 6$ ($p=4$ for $i=6$ and $p=0$ for $i=4$ or $5$) and $3\leq j\leq 6$; and
\eqn{deg6:K23:4 3 j}{
\begin{array}{*{20}l}
   C(\mu) &  \leq  & C(\mu\!-\!(w_6+w_5+6w_3+2\Delta w_6+\Delta w_j))\\
&&    + C(\mu\!-\!(w_6+w_5+3w_3+3\Delta w_6 \!-\!(w_{j+1}\!-\!w_j) )) ~~(3\leq j\leq 6).
\end{array}}

We analyze a special case in \refe{deg6:K23:4 i j} where $i=6$ and $j=6$.
For this case, we get the recurrence
\[
C(\mu)  \leq   C(\mu\!-\!(w_6+w_5+3(w_6\!-\!w_{4})+3\Delta w_6))+ C(\mu\!-\!(4w_6+w_5+ 5\Delta w_6   )).
\]
In the second branch we get the  graph $G^\dagger$, where $u'_4$ is a degree-7 vertex.
In the next step, either the degree-7 vertex is reduced by applying a reduction rule in Step~1
 or the algorithm will branch on the degree-7 vertex in Step~2.
%
For the former case, the measure will decrease by at least $2\Delta w_6$ by \refl{reduction_123}.
For the latter case, we can get $\sigma$ saved from the recurrence \refe{deg>=7:K24:i}.
We assume that $2\Delta w_6\geq \sigma$.
Then we can get following recurrence instead of the above one
\[
C(\mu)  \leq   C(\mu\!-\!(w_6+w_5+3(w_6\!-\!w_{4})+3\Delta w_6))+ C(\mu\!-\!(4w_6+w_5+ 5\Delta w_6 +\sigma )).
\]

Next, we further assume that there is no short edge $vv'$ with $\delta(v)=6$, $\delta(v')=5$ and $|N(v)\cap N(v')|=3$.
Then  the outer-degree of every degree-5 neighbor of a degree-6 vertex is at least 2.
\medskip

\noindent
Case (iii)  $d'=6$ and $k=3$:
Analogously with Case (ii), the first branch decreases the measure by at least
$2w_6 +\sum_{i=1,2,3}(w_{\delta(u_i)}-w_{\delta(u_i)-2})
+\sum_{x\in \{u_4,u_5,  u'_4, u'_5\}}\Delta w_{\delta(x)}+i^*w_3$.
We consider the second branch of deleting  $X=\{v,v',u_1,u_2,u_3\}$ from $G$
to construct $G^\dagger$ by adding edges $u_4u'_4$, $u_4u'_5$, $u_5u'_4$ and $u_5u'_5$, if necessary.
Let $p$   be the number of degree-5, 6 vertices in $\{u_1,u_2,u_3\}$, where each degree-5 or 6
 neighbor of $v$ has outer-degree at least 2.
Let $L$ denote the number of
edges in $G$ between $X$ and $V-X$  other than the four edges $vu_4$, $vu_5$, $v'u'_4$ and $v'u'_5$,
where $L\geq 3+p$.
For each vertex $x\in \{u_4, u_5,  u'_4, u'_5\}$,
  the degree of the vertex $x$ in $G-X$ is $\delta(x)-\ell_x-1$, where
 $\ell_x$ is the number of edges between $x$ and $\{u_1,u_2,u_3\}$.
Then
the weight change at $x$ from $G$ to $G^\dagger$ is at least
$$w_{\delta(x)}- w_{\delta(x)-\ell_x+1} \geq  -(w_{\delta(x)+1}-w_{\delta(x)}) + \ell_x \Delta w_6,$$
where $ \ell_x\Delta w_6$ is the lower bound on the weight decrease
caused by the deletion of the $\ell_x$ edges between $x$ and $\{u_1,u_2,u_3\}$.
Recall that $L\geq \sum_{x\in \{u_4, u_5,  u'_4, u'_5\}}\ell_x$.
Hence the measure decrease in the second branch is at least
$2w_6+\sum_{i=1,2,3} w_{\delta(u_i)} + L\Delta w_6
-\sum_{x\in \{u_4, u_5,  u'_4, u'_5\}}(w_{\delta(x)+1}-w_{\delta(x)})
 \geq
 2w_6+\sum_{i=1,2,3} w_{\delta(u_i)}  + (3+p)\Delta w_6
-\sum_{x\in \{u_4, u_5,  u'_4, u'_5\}}(w_{\delta(x)+1}-w_{\delta(x)})$.
By \refl{simplify_analysis}, we only need to consider the following  recurrences:
\eqn{deg6-6:K23:i>3}{\begin{array}{*{20}l}
C(\mu)&\leq& C(\mu \!-\!(2w_6 +3(w_i-w_{i-2}) +4\Delta w_j ) ) \\
&&  +C(\mu \!-\!(2w_6 +3w_i +(3+p)\Delta w_6-4(w_{j+1}\!-\!w_j) ) ),
\end{array}}
where $4\leq i\leq 6$  ($p=3$ for $i=5$ or $6$; and $p=0$ for $i=4$) and $3\leq j\leq 6$; and
%
\eqn{deg6-6:K23:i=3}{\begin{array}{*{20}l}
C(\mu) & \leq & C(\mu \!-\!(2w_6 \!+\!6 w_3  \!+\!4\Delta w_j ) )\\
&& +C(\mu \!-\!(2w_6  \!+\!3w_3 \!+\!3\Delta w_6 -4(w_{j+1}\!-\!w_j) ) ),
\end{array}}
where  $3\leq j\leq 6$.

We also analyze a special case in \refe{deg6-6:K23:i>3} where $i=6$ and $j=6$.
In the second branch we get the  graph $G^\dagger$ with four degree-7 vertices $u_4,u_5,u'_4$ and $u'_5$.
Analogously with the special case in Case (ii), in the second branch,
either the measure decreases by $2\Delta w_6\geq \sigma$ directly or we get shift $\sigma$ saved from \refe{deg>=7:K24:i}
 by branching on a degree-7 vertex.
 Then for this case we can get the following recurrence instead
\[
C(\mu)  \leq   C(\mu\!-\!(2w_6+3(w_6\!-\!w_{4})+4\Delta w_6))+ C(\mu\!-\!(5w_6+ 2\Delta w_6 +\sigma  )).
\]


From now on, we can assume that there is no short edges $vv'$ with $\delta(v)=\delta(v')=6$ and $|N(v)\cap N(v')|=3$.
Then the outer-degree of every degree-6 neighbor of a degree-6 vertex is at least 3.
\medskip

After Step~3 of ${\tt mis}6(G)$, for each degree-6 vertex $v$ in $G$, its degree-5 (resp., degree-6) neighbor is
of outer-degree $\geq 2$ (resp., $\geq 3$) at $v$, and it holds
\eqn{outer-edges}{ f_v\geq k_3+k_4+2k_5+3k_6.}

\subsection{Branching on vertices of maximum degree 6 in Step~4 of ${\tt mis}6(G)$} \label{sec:reduced-deg6}

We will derive recurrences for branchings in Step~4 of ${\tt mis}6(G)$.
After Step~3, we can assume that the current graph $G$ is a reduced graph with maximum degree 6
such that there is no short edge.
Let $v$ be an optimal vertex $v$ of degree 6 selected in Step~4 of ${\tt mis}6(G)$.

We define
\eqn{set-lambda}{
 \lambda_6(k_3,k_4,k_5,k_6)
=\left\{ \begin{array}{cl}
 \min \{(12+k_6)\Delta w_6, w_3+6\Delta w_6\}  & \mbox{if  $k_6\leq 3$ and $k_3\!+\!k_4\geq 2$ }\\
 (6+k_5+2k_6)\Delta w_6  & \mbox{if  $k_6\leq 3$ and $k_3\!+\!k_4\leq 1$ }\\
 (6+k_5+2k_6)\Delta w_6  & \mbox{if  $k_6=4$ and $k_3\!+\!k_4\geq 1$ }\\
 17\Delta w_6  & \mbox{if  $k_6=4$ and $k_5=2$ }\\
 (16+2k_4+3k_5)\Delta w_6 & \mbox{if  $k_6=5$ }\\
  22\Delta w_6 & \mbox{if $k_6=6$. }
  \end{array}
\right.}

Then we have:
\lem{decrease-2nd-branch}{Let $v$ be an optimal degree-6 vertex in Step~4 of ${\tt mis}6(G)$.
Then $\Delta (\overline{N[v]}) \geq \lambda_6(k_3,k_4,k_5,k_6)$. }

\pf{By \refe{outer-edges}, we have
$\Delta (\overline{N[v]}) \geq  f_v\Delta w_6   \geq (k_3+k_4+2k_5+3k_6)\Delta w_6=(6+k_5+2k_6)\Delta w_6$.
This proves the cases of ``$k_6\leq 3$ and $k_3\!+\!k_4\leq 1$,''   ``$k_6=4$ and $k_3\!+\!k_4\geq 1$,''
and ``$k_6=5$ and $k_3=1$'' (the case of ``$k_6=5$ and $k_3\neq1$'' will be treated next).

By \refl{degree-change} and the definition of optimal vertices imply
the case of  ``$k_6=4$ and $k_5=2$'', ``$k_6=5$ and $k_3\neq1$'' and ``$k_6=6$.''

Finally we show the case  of  ``$k_6\leq 3$ and $k_3\!+\!k_4\geq 2$.''
If $N^*(v)=\emptyset$ then each degree-3,4 neighbor of $v$ also has at least two neighbors
in $N_2(v)$, and we again obtain
$\Delta (\overline{N[v]}) \geq  f_v\Delta w_6   \geq (12+k_6)\Delta w_6$.
If $N^*(v)\neq \emptyset$, then \refl{outside_decrease}(ii) implies that
  $\Delta(\overline{N[v]}) \geq
  \min \{2w_3 , w_3+ 2(\delta(v)-3)\Delta w_6\} \geq w_3\!+\!6\Delta w_6$.
This proves the case of  ``$k_6\leq 3$ and $k_3\!+\!k_4\geq 2$.''
}\medskip

By \refl{decrease-2nd-branch}, we obtain the recurrence \refe{max-deg} for $d=6$ as follows:
\eqn{deg=6}{\begin{array}{*{20}l}
C(\mu) &\leq &  C(\mu-(w_6+ k_3 w_3+ k_4(w_4-w_3)+ k_5(w_5\! -\! w_4) + k_6(w_6\! -\! w_5) ))\\
&& + C(\mu-(w_6+ k_3 w_3+ k_4 w_4+ k_5 w_5  +k_6 w_6 +\lambda_6(k_3,k_4,k_5,k_6)   ))
\end{array}}
for all nonnegative integers $(k_3,k_4,k_5,k_6)$ with  $k_3+k_4+k_5+k_6=6$.

\subsection{Finial step}\label{sec:max-degree<6}

We have derived recurrences for all branching operations  in algorithm ${\tt mis}6(G)$
except for Step~5 which invokes the fast algorithms for MIS-$5$ in \cite{XN:5MIS}.
To  determine the largest branching factor to algorithm ${\tt mis}6(G)$
 using our divide-and-conquer method in Section~\ref{sec_dc},
we  combine all the above recurrences with the weight setting used
to determine the branching factor to algorithms for MIS-$5$  in \cite{XN:5MIS}.

The algorithm for MIS-$5$ in \cite{XN:5MIS}  runs in
 $1.17366^{\mu_5(G)} \mu_5(G)^{O(1)} $ time for a degree-5 graph $G$
with measure $\mu_5(G)=\sum_{1\leq i\leq 5}w^{\langle 5 \rangle}_i n_i$
where $n_i$ is the number of degree-$i$ vertices in $G$, and
$w^{\langle 5 \rangle}_i$ is the weight of a degree-$i$ vertex such that
$w^{\langle 5 \rangle}_0=w^{\langle 5 \rangle}_1=w^{\langle 5 \rangle}_2=0$,
$w^{\langle 5 \rangle}_3=0.50907$,
$w^{\langle 5 \rangle}_4=0.82427$ and $w^{\langle 5 \rangle}_5=1$.
Based on \refl{DCalg}, we include  the following three constraints into the current set of recurrences.
\eqn{MIS5}{\begin{array}{*{20}l}
 C(\mu)\!\leq \! 1.17366^{\frac{w^{\langle 5 \rangle}_3}{w_3}\mu}, &  ~~
     C(\mu)\!\leq \! 1.17366^{\frac{w^{\langle 5 \rangle}_4}{w_4}\mu} , ~~\mbox{and}\\
 C(\mu)\!\leq \! 1.17366^{\frac{w^{\langle 5 \rangle}_5}{w_5}\mu}. &
\end{array}}
\invis{
  c(1)= 1.17366^(1/x(5)) -x(1); 
  c(2)= 1.17366^(0.82427/x(4)) -x(1); 
  c(3)= 1.17366^(0.50907/x(3)) -x(1); 
}%


Recurrences
\refe{deg>=7:K24:i}
to
\refe{deg6-6:K23:i=3} and \refe{deg=6} together with \refe{MIS5}
generate the constraints in a quasiconvex program to minimize the largest branching factor $\tau$.
By solving the quasiconvex program
according to the method introduced in~\cite{eQP},
we get an upper bound $1.18922$ on the branching factor for all recurrences by setting vertex weights    as
\eqn{weight_setting}{
 w_i
=\left\{ \begin{array}{cl}
 0 & \mbox{for  $i=0,1$ and $2$}\\
 0.49969  & \mbox{for $i=3$}\\
 0.76163  & \mbox{for $i=4$}\\
 0.92401  & \mbox{for $i=5$}\\
 1  & \mbox{for $i=6$}\\
 w_{6} + (i-6)(w_6\!-\!w_5) & \mbox{for $i\geq 7$.}
  \end{array}
\right.}
Now a feasible value of the shift $\sigma$ is $0.10647$.
This verifies \refl{branch-general} with $\theta=6$.


\section{Analysis of ${\tt mis}7(G)$} \label{app_deg7}

In the same manner of  Section~\ref{sec:measure},
we analyze the largest branching factor of recurrences for the branchings in ${\tt mis}7(G)$.
All notations except for a new vertex weight in ${\tt mis}7(G)$  stand for the same meaning in
Section~\ref{sec:measure}.

Recall that, for MIS-7,
 we assume that $w_0=w_1=w_2=0\leq w_3\leq w_4\leq w_5\leq w_6\leq w_7= 1\leq w_8\leq \cdots , $
and the values of $w_3$, $w_4$, $w_5$  and $w_6$ will be determined
after we analyze how the measure changes after each step of the algorithm.

%
To simplify our analysis, we assume that
\eqn{7new-1}
{18\Delta w_7\leq w_3.}
\invis{
 c(6)=18*(x(7)-x(6))- x(3) ;  
 }
%
\invis{
 c( )= 2*(x(7)-x(6)) -(x(6)-x(5)); 
 c( )= (x(6)-x(5)) -(x(5)-x(4)); 
 c( )= (x(5)-x(4)) -(x(4)-x(3)) ; 
 c( )= (x(4)-x(3)) -x(3) ; 
}
We impose the next constraint so that the measure does not increase after contracting vertices.
\eqn{7weight-merge}{w_i+w_j\geq w_{i+j-2},~~~ 3\leq i, j\leq 7.}
\invis{
 c( )=x(4) -x(3)-x(3) ; 
 c( )= x(5)-x(3)-x(4)   ; 
 c( )= x(6)-x(4)-x(4) ; 
 c( )= x(6)-x(5)-x(3)   ; 
 c( )= x(6)+(x(6)-x(5))-x(5)-x(4)  ; 
 c( )= x(6)+2*(x(6)-x(5))-x(5)-x(5)  ; 
 %
 c( )= x(7)-x(6)-x(3)   ; 
 c( )= x(7)+(x(7)-x(6))-x(5)-x(4)  ; 
 c( )= x(7)+2*(x(7)-x(6))-x(6)-x(5)  ; 
 c( )= x(7)+3*(x(7)-x(6))-x(6)-x(6)  ; 
 }%

We can see that \refl{weight-sum} still holds in ${\tt mis}7(G)$
and then the measure of the graph will not increase after Step~1 of ${\tt mis}7(G)$.
Next we derive a recurrence of each branching step of ${\tt mis}7(G)$.

\medskip
\noindent
Step~1: It is easy to see that the statement of \refl{outside_decrease} still holds for $\theta=7$
   even after replacing   `$\Delta w_6$' with `$\Delta w_7$' in it.
Based on the $\theta=7$ version of \refl{outside_decrease},
we see that  every vertex $v$ with $\delta(v)\geq 8$ satisfies
 $$\Delta (\overline{N[v]})\geq   \min \{ 2\delta(v)\Delta w_7,  2w_3 , w_3+ 2(\delta(v)-3)\} \geq   14\Delta w_7$$
  by  \refe{new-min-delta} and \refe{7new-1}.

\medskip
\noindent
Step~2: We use recurrences \refe{max-deg} for  branching on a vertex of degree $\geq 8$ in Step~2 of ${\tt mis}7(G)$.
By \refl{simplify_analysis}, we only have to consider the case of $d=8$ for the recurrences with $k_i=8$ $(3\leq i \leq 8)$.
Thus we get  recurrences
\eqn{deg>=8:K24:i}{
 \begin{array}{*{20}l}
C(\mu) &\leq &  C(\mu\!-\!(w_d+\sum_{i=3}^d k_i \Delta w_i)
+ C(\mu\!-\!(w_d +\sum_{i=3}^d k_i w_i + \Delta (\overline{N[v]}))) \\
&\leq &   \max_{3\leq i\leq 8} [C(\mu\!-\!(w_8\!+\!8\Delta w_i)) + C(\mu\!-\!(w_8 \!+\!8 w_i\!+\!  14\Delta w_7  ))].
\end{array}
}
\invis{
 c( )=-1+x(1)^(-(x(8)+8*(x(7)-x(6)) )) +x(1)^(-( x(8)+8*x(7)  +14*(x(7)-x(6)) )) ;
 c( )=-1+x(1)^(-(x(8)+8*(x(6)-x(5)) )) +x(1)^(-( x(8)+8*x(6)  +14*(x(7)-x(6)) )) ;
 c( )=-1+x(1)^(-(x(8)+8*(x(5)-x(4)) )) +x(1)^(-( x(8)+8*x(5)  +14*(x(7)-x(6)) )) ;
 c( )=-1+x(1)^(-(x(8)+8*(x(4)-x(3)) )) +x(1)^(-( x(8)+8*x(4)  +14*(x(7)-x(6)) )) ;
 c( )=-1+x(1)^(-(x(8)+8*(x(3)     ) )) +x(1)^(-( x(8)+8*x(3)  +14*(x(7)-x(6)) )) ;
}%

\medskip
\noindent
Step~3: We consider branching on an optimal short edge  $vv'$ in Step~3 of ${\tt mis}7(G)$.
We see that $|N(v)\cap N(v')|\geq 6$ cannot occur otherwise $v$ would dominate  $v'$.

\noindent
(i) We derive recurrences in case of $|N(v)\cap N(v')|=5$.
Analogously with Case (i) in Section~\ref{sec:bipartite}, we get recurrences
\eqn{deg7:K25:i}{
\begin{array}{*{20}l}
   C(\mu) &  \leq  & C(\mu\!-\!(2w_7+5(w_i\!-\!w_{i-2}) +2\Delta w_7))\\
&&    + C(\mu\!-\!(2w_7+5w_i+4\Delta w_7 )) ~~(i=4,5,6,7);
\end{array}}
and
\eqn{deg7:K24:3}{
   C(\mu)  \leq   C(\mu\!-\!(2w_7+10w_3 +2\Delta w_7))  + C(\mu\!-\!(2w_7+5w_3+4\Delta w_7 )). }
\invis{
c(55)= -1+x(1)^(-(2*x(7)+5*(x(7)-x(5))+2*(x(7)-x(6)) )) + x(1)^(-(2*x(7)+5*x(7)+4*(x(7)-x(6)) ));
c(55)= -1+x(1)^(-(2*x(7)+5*(x(6)-x(4))+2*(x(7)-x(6)) )) + x(1)^(-(2*x(7)+5*x(6)+4*(x(7)-x(6)) ));
c(56)= -1+x(1)^(-(2*x(7)+5*(x(5)-x(3))+2*(x(7)-x(6)) )) + x(1)^(-(2*x(6)+5*x(5)+4*(x(7)-x(6)) ));
c(57)= -1+x(1)^(-(2*x(7)+5*x(4)         +2*(x(7)-x(6)) )) + x(1)^(-(2*x(6)+5*x(4)+4*(x(7)-x(6)) ));
c(58)= -1+x(1)^(-(2*x(7)+10*x(3)         +2*(x(7)-x(6)) )) + x(1)^(-(2*x(6)+5*x(3)+4*(x(7)-x(6)) ));
}

\noindent
(ii) We derive recurrences in case of $|N(v)\cap N(v')|=4$.
Denote  $X=N(v)\cap N(v')=\{u_i\mid i=1,2,3,4\}$,
 $N(v)-N[v']=\{u_5, u_6\}$ and $N(v')-N[v]=\{u'_5, u'_6\}$.
Let $i^*$ denote the number of degree-3 vertices $u\in \{u_1,u_2,u_3,u_4\}$,
where we assume that for each $i\leq i^*$, $u_i$ is a degree-3 neighbor of $v$.
Analogously with  Case (iii) in Section~\ref{sec:bipartite}, the first branch of deleting vertices $v$ and $v'$ decreases
the measure by at least
$2w_7 +\sum_{i=1,2,3,4}(w_{\delta(u_i)}-w_{\delta(u_i)-2})
+\sum_{x\in \{u_5,u_6,  u'_5, u'_5\}}\Delta w_{\delta(x)}+i^*w_3$.
In the second branch of deleting  $X=\{v,v',u_1,u_2,u_3, u_4\}$ from $G$
to construct $G^\dagger$ by adding edges $u_5u'_5$, $u_5u'_6$, $u_6u'_5$ and $u_6u'_6$, if necessary.
Let $p$ be the number of degree-7 vertices in $\{u_1,u_2,u_3,u_4\}$, where each degree-7
 neighbor of $v$ has outer-degree at least 2 (otherwise there would be a short edge  $aa'$ with $|N(a)\cap N(a')|\geq5$).
Let $L$ denote the number of
edges in $G$ between $X$ and $V-X$  other than the four edges $vu_5$, $vu_6$, $v'u'_5$ and $v'u'_6$,
where $L\geq 4+p$.
The following analysis is the same as Case (iii) in Section~\ref{sec:bipartite}.
The decrease of the measure in the second branch is at least
$2w_7+\sum_{i=1,2,3,4} w_{\delta(u_i)}  + (4+p)\Delta w_6
-\sum_{x\in \{u_5, u_6,  u'_5, u'_6\}}(w_{\delta(x)+1}-w_{\delta(x)})$.
By \refl{simplify_analysis}, we only need to consider the following  recurrences:
\eqn{deg7-7:K24:i>3}{\begin{array}{*{20}l}
C(\mu)&\leq& C(\mu \!-\!(2w_7 +4(w_i-w_{i-2}) +4\Delta w_j ) ) \\
&&  +C(\mu \!-\!(2w_7 +4w_i +(4+p)\Delta w_7-4(w_{j+1}\!-\!w_j) ) ),
\end{array}}
where $4\leq i\leq 7$  ($p=4$ for $i=7$;  and $p=0$ for $i=4,5$ and $6$) and $3\leq j\leq 7$; and
%
\eqn{deg7-7:K24:i=3}{\begin{array}{*{20}l}
C(\mu) & \leq & C(\mu \!-\!(2w_7 \!+\!8 w_3  \!+\!4\Delta w_j ) )\\
&& \!+\!C(\mu \!-\!(2w_7  \!+\!4w_3 \!+\!4\Delta w_7 -4(w_{j+1}\!-\!w_j) ) ),
\end{array}}
where  $3\leq j\leq 7$.
\medskip

\medskip
\noindent
Step~4: We again use recurrences \refe{max-deg} for branching on an optimal vertex  $v$ in Step~4 of ${\tt mis}7(G)$.
Now the graph has no short edge, and every degree-7 vertex has outer-degree at least 3 at a degree-7 neighbor of it.
Hence it holds
\eqn{outer-edges7}{ f_v\geq k_3+k_4+k_5+k_6+3k_7.}
We define
\eqn{7set-lambda}{
 \lambda_7(k_7)
=\left\{ \begin{array}{cl}
 (14\!+\!k_7)\Delta w_7  & \mbox{if $k_7\leq 5$}\\
  (22-2k_3-k_4)\Delta w_7 & \mbox{if $k_7= 6$}\\
  26\Delta w_7 & \mbox{if $k_7= 7$}.
  \end{array}
\right.}
Then we have:
\lem{7decrease-2nd-branch}{Let $v$ be an optimal degree-7 vertex in Step~4 of ${\tt mis}7(G)$.
Then $\Delta (\overline{N[v]}) \geq \lambda_7(k_7)$. }

\pf{
First consider the case of $N^*(v)=\emptyset$.
Then each neighbor of $v$ has at least two neighbors
in $N_2(v)$ and each degree-7 neighbor of $v$ has at least three neighbors in $N_2(v)$.
By \refe{outer-edges7}, we obtain
$\Delta (\overline{N[v]}) \geq  f_v\Delta w_7   \geq (14+k_7)\Delta w_7$ for $k_7\leq 5$.
By \refl{degree-change}, it holds
$\Delta (\overline{N[v]}) \geq (f_v+(f_v \! -\! |N_2(v)|))\Delta w_7$.
For  $k_7=6$ (resp., $k_7=7$), this and  the definition of optimal vertices
imply that
  $\Delta (\overline{N[v]}) \geq (f_v+(f_v \! -\! |N_2(v)|)) \Delta w_7 \geq (22-2k_3-k_4)\Delta w_7$
(resp., $\geq  26\Delta w_7$).




If $N^*(v)\neq \emptyset$, then the $\theta=7$ version of \refl{outside_decrease}(ii)  implies that
  $\Delta(\overline{N[v]}) \geq
  \min \{ 2w_3 , w_3+ 2(\delta(v)-3)\Delta w_7\} \geq w_3\!+\!8\Delta w_7$,
which is larger than any of  $(14+k_7)\Delta w_7$, $(22-2k_3-k_4)\Delta w_7$ and $26 \Delta w_7$  by \refe{7new-1}.
This proves all the cases.
}\medskip

By \refl{decrease-2nd-branch}, we obtain the recurrence \refe{max-deg} for $d=7$ as follows.

\noindent
Case 1 ($k_7\leq 5$):
\eqn{deg=7-1}{\begin{array}{*{20}l}
C(\mu) &\leq &  C(\mu-(w_7+ 7(w_i\! -\! w_{i-1}) )) + C(\mu-(w_7+ 7w_i +14 \Delta w_7  )),
\end{array}}
where $3\leq i\leq 6$.
\invis{
c( )= -1+x(1)^(-(x(7)+ 0*(x(7)-x(6))+7*(x(6)-x(5))+0*(x(5)-x(4))+0*(x(4)-x(3))+0*x(3) ))  +  x(1)^(-(x(7) +0*x(7) +7*x(6) +0*x(5)+0*x(4)+0*x(3) +14*(x(7)-x(6)) ));
c( )= -1+x(1)^(-(x(7)+ 0*(x(7)-x(6))+0*(x(6)-x(5))+7*(x(5)-x(4))+0*(x(4)-x(3))+0*x(3) ))  +  x(1)^(-(x(7) +0*x(7) +0*x(6) +7*x(5)+0*x(4)+0*x(3) +14*(x(7)-x(6)) ));
c( )= -1+x(1)^(-(x(7)+ 0*(x(7)-x(6))+0*(x(6)-x(5))+0*(x(5)-x(4))+7*(x(4)-x(3))+0*x(3) ))  +  x(1)^(-(x(7) +0*x(7) +0*x(6) +0*x(5)+7*x(4)+0*x(3) +14*(x(7)-x(6)) ));
c( )= -1+x(1)^(-(x(7)+ 0*(x(7)-x(6))+0*(x(6)-x(5))+0*(x(5)-x(4))+0*(x(4)-x(3))+7*x(3) ))  +  x(1)^(-(x(7) +0*x(7) +0*x(6) +0*x(5)+0*x(4)+7*x(3) +14*(x(7)-x(6)) ));
}

\noindent
Case 2 ($k_7=6$ and $k_3=1$):
\eqn{deg=7-2-1}{\begin{array}{*{20}l}
C(\mu) &\leq &  C(\mu-(w_7+ 6(w_7\! -\! w_6)+w_3 )) + C(\mu-(7w_7+w_3 +20 \Delta w_7  )).
\end{array}}
\invis{
c( )= -1+x(1)^(-(x(7)+ 6*(x(7)-x(6))+0*(x(6)-x(5))+0*(x(5)-x(4))+0*(x(4)-x(3))+1*x(3) ))  +  x(1)^(-(x(7) +6*x(7) +0*x(6) +0*x(5)+0*x(4)+1*x(3) +20*(x(7)-x(6)) ));
}

\noindent
Case 3 ($k_7=6$ and $k_4 = 1$):
\eqn{deg=7-2-2}{\begin{array}{*{20}l}
C(\mu) &\leq &  C(\mu-(w_7+ 6(w_7\! -\! w_6)+(w_4-w_3) )) \\
&&+ C(\mu-(7w_7+w_4 +21 \Delta w_7  )).
\end{array}}
\invis{
c( )= -1+x(1)^(-(x(7)+ 6*(x(7)-x(6))+0*(x(6)-x(5))+0*(x(5)-x(4))+1*(x(4)-x(3))+0*x(3) ))  +  x(1)^(-(7*x(7) +0*x(6) +0*x(5)+1*x(4)+0*x(3) +21*(x(7)-x(6)) ));  
}

\noindent
Case 4 ($k_7=6$ and $k_5+k_6=1$):
\eqn{deg=7-2-3}{\begin{array}{*{20}l}
C(\mu) &\leq &  C(\mu-(w_7+ 6(w_7\! -\! w_6)+(w_i\! -\! w_{i-1}) )) \\
&& + C(\mu-(7w_7+w_i +22 \Delta w_7  )),
\end{array}}
where $5\leq i\leq 6$.
\invis{
c( )= -1+x(1)^(-(x(7)+ 6*(x(7)-x(6))+1*(x(6)-x(5))+0*(x(5)-x(4))+0*(x(4)-x(3))+0*x(3) ))  +  x(1)^(-(x(7) +6*x(7) +1*x(6) +0*x(5)+0*x(4)+0*x(3) +22*(x(7)-x(6)) ));
c( )= -1+x(1)^(-(x(7)+ 6*(x(7)-x(6))+0*(x(6)-x(5))+1*(x(5)-x(4))+0*(x(4)-x(3))+0*x(3) ))  +  x(1)^(-(x(7) +6*x(7) +0*x(6) +1*x(5)+0*x(4)+0*x(3) +22*(x(7)-x(6)) ));
}

\smallskip
\noindent
Case 5 ($k_7=7$):
\eqn{deg=7-3}{\begin{array}{*{20}l}
C(\mu) &\leq &  C(\mu-(w_7+ 7(w_7\! -\! w_6) )) + C(\mu-(8w_7+26 \Delta w_7  )).
\end{array}}
\invis{
c( )= -1+x(1)^(-(x(7)+ 7*(x(7)-x(6))+0*(x(6)-x(5))+0*(x(5)-x(4))+0*(x(4)-x(3))+0*x(3) ))  +  x(1)^(-(x(7) +7*x(7) +0*x(6) +0*x(5)+0*x(4)+0*x(3) +26*(x(7)-x(6)) ));
}

\medskip

We have derived recurrences for all branching operations  in algorithm ${\tt mis}7(G)$
except for Step~5 which invokes  algorithms ${\tt mis}6(G)$.
To  determine the largest branching factor to algorithm ${\tt mis}7(G)$ analogously with the previous section,
we  combine all the above recurrences with the weight setting used for ${\tt mis}6(G)$.
By \refl{DCalg}, we  include  the following four constraints into the current set of recurrences.
\eqn{MIS6_solution}{\begin{array}{*{20}l}
 C(\mu)\!\leq \! 1.18922^{\frac{w^{\langle 6 \rangle}_3}{w_3}\mu}  , &   ~~   C(\mu)\!\leq \! 1.18922^{\frac{w^{\langle 6 \rangle}_4}{w_4}\mu}  , \\
 C(\mu)\!\leq \! 1.18922^{\frac{w^{\langle 6 \rangle}_5}{w_5}\mu}  ,&\mbox{and}~~C(\mu)\!\leq \! 1.18922^{\frac{w^{\langle 6 \rangle}_6}{w_6}\mu}  ,
\end{array}}
where $w^{\langle 6 \rangle}_3=0.49969$,
$w^{\langle 6 \rangle}_4=0.76163$, $w^{\langle 6 \rangle}_5=0.92401$ and $w^{\langle 6 \rangle}_6=1$.
\invis{
  c(1)= 1.18922^(1/x(6)) -x(1); 
  c(2)= 1.18922^(0.92401/x(5)) -x(1); 
  c(3)= 1.18922^(0.76163/x(4)) -x(1); 
  c(4)= 1.18922^(0.49969/x(3)) -x(1); 
}%

The above assumptions and recurrences together with \refe{MIS6_solution}
generate the constraints in our quasiconvex program.
By solving the quasiconvex program,
we get an upper bound $1.19698$ on the branching factor for all recurrences by setting vertex weights as
\eqn{7weight_setting}{
 w_i
=\left\{ \begin{array}{cl}
 0 & \mbox{for  $i=0,1$ and $2$}\\
 0.65077  & \mbox{for $i=3$}\\
 0.78229  & \mbox{for $i=4$}\\
 0.89060 & \mbox{for $i=5$}\\
 0.96384 & \mbox{for $i=6$}\\
 1  & \mbox{for $i=7$}\\
 w_{7} + (i-7)(w_7\!-\!w_6) & \mbox{for $i\geq 8$.}
  \end{array}
\right.}
This verifies \refl{branch-general} with $\theta=7$.


\section{Analysis of ${\tt mis}8(G)$} \label{app_deg8}

Recall that for MIS-8,
we assume
 $w_0=w_1=w_2=0\leq w_3\leq w_4\leq w_5\leq w_6\leq w_7\leq w_8= 1\leq w_9\leq \cdots , $
and the values of $w_3$, $w_4$, $w_5$, $w_6$  and $w_7$ will be determined
after we analyze how the measure changes after each step of the algorithm.

To simplify analysis, we assume that
\eqn{8new-1}
{26\Delta w_8\leq w_3.}
\invis{
 c(6)=26*(x(8)-x(7))- x(3) ;  
 }
%
\invis{
 c( )= 2*(x(8)-x(7)) -(x(7)-x(6)); 
 c( )= x(7)-x(6) -(x(6)-x(5)); 
 c( )= (x(6)-x(5)) -(x(5)-x(4)); 
 c( )= (x(5)-x(4)) -(x(4)-x(3)) ; 
 c( )= (x(4)-x(3)) -x(3) ; 
}
To ensure that contracting vertices never increase the measure, we impose
the next constraint.
\eqn{8weight-merge}{w_i+w_j\geq w_{i+j-2},~~~ 3\leq i, j\leq 8.}
\invis{
 c(15)=x(4) -x(3)-x(3) ; 
 c(16)= x(5)-x(3)-x(4)   ; 
 c(17)= x(6)-x(4)-x(4) ; 
 c(18)= x(6)-x(5)-x(3)   ; 
 c(19)= x(6)+(x(6)-x(5))-x(5)-x(4)  ; 
 c(20)= x(6)+2*(x(6)-x(5))-x(5)-x(5)  ; 
 %
 c(21)= x(7)-x(6)-x(3)   ; 
 c(22)= x(7)+(x(7)-x(6))-x(6)-x(4)  ; 
 c(23)= x(7)+2*(x(7)-x(6))-x(6)-x(5)  ; 
 c(24)= x(7)+3*(x(7)-x(6))-x(6)-x(6)  ; 
 %
 c(25)= x(8)-x(7)-x(3)   ; 
 c(26)= x(8)+(x(8)-x(7))-x(7)-x(4)  ; 
 c(27)= x(8)+2*(x(8)-x(7))-x(7)-x(5)  ; 
 c(28)= x(8)+3*(x(8)-x(7))-x(7)-x(6)  ; 
 c(29)= x(8)+4*(x(8)-x(7))-x(7)-x(7)  ; 
 }%

\medskip

Next we analyze each step of the algorithm.

\noindent
Step~1:  We can see that \refl{weight-sum} still holds in ${\tt mis}8(G)$
and then the measure of the graph will not increase after Step~1 of ${\tt mis}8(G)$.

\medskip
\noindent
Step~2:  Using recurrences \refe{max-deg} for  branching on a vertex of degree $\geq 9$ in Step~2,
  we get recurrences:
\eqn{deg>=9:K24:i}{
 \begin{array}{*{20}l}
C(\mu) &\leq &  C(\mu\!-\!(w_d+\sum_{i=3}^d k_i \Delta w_i)
+ C(\mu\!-\!(w_d +\sum_{i=3}^d k_i w_i + \Delta (\overline{N[v]}))) \\
&\leq &   \max_{3\leq i\leq 9} [C(\mu\!-\!(w_9\!+\!9\Delta w_i)) + C(\mu\!-\!(w_9 \!+\!9 w_i\!+\!  16\Delta w_8  ))].
\end{array}
}
\invis{
 c( )=-1+x(1)^(-(x(9)+9*(x(8)-x(7)) )) +x(1)^(-( x(9)+9*x(8)  +16*(x(8)-x(7)) )) ;
 c( )=-1+x(1)^(-(x(9)+9*(x(7)-x(6)) )) +x(1)^(-( x(9)+9*x(7)  +16*(x(8)-x(7)) )) ;
 c( )=-1+x(1)^(-(x(9)+9*(x(6)-x(5)) )) +x(1)^(-( x(9)+9*x(6)  +16*(x(8)-x(7)) )) ;
 c( )=-1+x(1)^(-(x(9)+9*(x(5)-x(4)) )) +x(1)^(-( x(9)+9*x(5)  +16*(x(8)-x(7)) )) ;
 c( )=-1+x(1)^(-(x(9)+9*(x(4)-x(3)) )) +x(1)^(-( x(9)+9*x(4)  +16*(x(8)-x(7)) )) ;
 c( )=-1+x(1)^(-(x(9)+9*(x(3)     ) )) +x(1)^(-( x(9)+9*x(3)  +16*(x(8)-x(7)) )) ;
}%

\medskip
\noindent
Step~3:
 We consider branching on an optimal short edge  $vv'$ in Step~3.
We see that $|N(v)\cap N(v')|\geq 7$ cannot occur, otherwise $v$ would dominate  $v'$.

\noindent
(i)
For the case of $|N(v)\cap N(v')|=6$,
 we get the following recurrences for the branch (analogously with Case (i) in Section~\ref{sec:bipartite})
\eqn{deg8:K26:i}{
\begin{array}{*{20}l}
   C(\mu) &  \leq  & C(\mu\!-\!(2w_8+6(w_i\!-\!w_{i-2}) +2\Delta w_8))\\
&&    + C(\mu\!-\!(2w_8+6w_i+2\Delta w_8 )) ~~(i=3,4,5,6,7).
\end{array}}
\invis{
c(55)= -1+x(1)^(-(2*x(8)+6*(x(8)-x(6))+2*(x(8)-x(7)) )) + x(1)^(-(2*x(8)+6*x(8)+4*(x(8)-x(7)) ));
c(55)= -1+x(1)^(-(2*x(8)+6*(x(7)-x(5))+2*(x(8)-x(7)) )) + x(1)^(-(2*x(8)+6*x(7)+4*(x(8)-x(7)) ));
c(55)= -1+x(1)^(-(2*x(8)+6*(x(6)-x(4))+2*(x(8)-x(7)) )) + x(1)^(-(2*x(8)+6*x(6)+4*(x(8)-x(7)) ));
c(56)= -1+x(1)^(-(2*x(8)+6*(x(5)-x(3))+2*(x(8)-x(7)) )) + x(1)^(-(2*x(8)+6*x(5)+4*(x(8)-x(7)) ));
c(57)= -1+x(1)^(-(2*x(8)+6*x(4)         +2*(x(8)-x(7)) )) + x(1)^(-(2*x(8)+6*x(4)+4*(x(8)-x(7)) ));
c(58)= -1+x(1)^(-(2*x(8)+6*x(3)         +2*(x(8)-x(7)) )) + x(1)^(-(2*x(8)+6*x(3)+4*(x(8)-x(7)) ));
}

\noindent
(ii) For the case of $|N(v)\cap N(v')|=5$,
 we get the following recurrences
\eqn{deg8-8:K25:i>3}{\begin{array}{*{20}l}
C(\mu)&\leq& C(\mu \!-\!(2w_8 +5(w_i-w_{i-2}) +4\Delta w_j ) ) \\
&&  +C(\mu \!-\!(2w_8 +5w_i +(5+p)\Delta w_8-4(w_{j+1}\!-\!w_j) ) ),
\end{array}}
where $4\leq i\leq 8$  ($p=5$ for $i=8$ and $p=0$ for $i=4,5,6$ and $7$) and $3\leq j\leq 8$; and
%
\eqn{deg8-8:K25:i=3}{\begin{array}{*{20}l}
C(\mu) & \leq & C(\mu \!-\!(2w_8 \!+\!10 w_3  \!+\!4\Delta w_j ) ) \\
&&+C(\mu \!-\!(2w_8  \!+\!5w_3 \!+\!4\Delta w_7 -4(w_{j+1}\!-\!w_j) ) ),
\end{array}}
where  $3\leq j\leq 8$.

\noindent
(iii)
For the case of $|N(v)\cap N(v')|=4$,  we get the following recurrences
\eqn{deg8-8:K34:i>3}{\begin{array}{*{20}l}
C(\mu)&\leq& C(\mu \!-\!(2w_8 +4(w_i-w_{i-2}) +6\Delta w_j ) ) \\
&&  +C(\mu \!-\!(2w_8 +4w_i +(4+p)\Delta w_8-6(w_{j+1}\!-\!w_j) ) ),
\end{array}}
where $4\leq i\leq 8$  ($p=8$ for $i=8$; $p=4$ for $i=7$; and $p=0$ for $i=4,5$ and $6$) and $3\leq j\leq 8$; and
%
\eqn{deg8-8:K34:i=3}{\begin{array}{*{20}l}
C(\mu) & \leq & C(\mu \!-\!(2w_8 \!+\!8 w_3  \!+\!6\Delta w_j ) ) \\
&&+C(\mu \!-\!(2w_8  \!+\!4w_3 \!+\!4\Delta w_7 -6(w_{j+1}\!-\!w_j) ) ),
\end{array}}
where  $3\leq j\leq 8$.


\medskip

\medskip
\noindent
Step~4: Using recurrences \refe{max-deg} for branching on an optimal vertex of degree 8 in Step~4, we can get recurrences:
\eqn{deg=8}{\begin{array}{*{20}l}
C(\mu) &\leq &  C(\mu-(w_8+ \sum_{i=3}^8 k_i \Delta w_i )) + C(\mu-(w_8+ \sum_{i=3}^8 k_i w_i +\lambda_8(k_8)   )),
\end{array}}
for all nonnegative integers $(k_3,k_4,k_5,k_6,k_7,k_8)$ with  $k_3+k_4+k_5+k_6+k_7+k_8=8$ and
\eqn{8set-lambda}{
 \lambda_8(k_8)
=\left\{ \begin{array}{cl}
 (16\!+\! 2k_8)\Delta w_8  & \mbox{if $k_8\leq 7$}\\
    36\Delta w_8 & \mbox{if $k_8=8$}.
  \end{array}
\right.}

The correctness of the above recurrences relies on the following lemma, which corresponds to \refl{7decrease-2nd-branch}.

\lem{8decrease-2nd-branch}{Let $v$ be an optimal degree-8 vertex in Step~4 of ${\tt mis}8(G)$.
Then $\Delta (\overline{N[v]}) \geq \lambda_8(k_8)$. }

\pf{If $N^*(v)=\emptyset$. Then each neighbor of $v$ has at least two neighbors
in $N_2(v)$ and each degree-8 neighbor of $v$ has at least four neighbors in $N_2(v)$.
We obtain
$\Delta (\overline{N[v]}) \geq  f_v\Delta w_8   \geq (16+2k_8)\Delta w_8$.
By the definition of optimal vertices, we know that
when $k_8=8$, it holds $f_v\geq 36$ and then $\Delta (\overline{N[v]}) \geq
36\Delta w_8$.

We see thatthe statement of \refl{outside_decrease}(ii) still holds for $\theta=8$
   even after replacing   `$\Delta w_6$' with `$\Delta w_8$' in it.
This implies that if $N^*(v)\neq \emptyset$, then
  $\Delta(\overline{N[v]}) \geq
  \min \{ 2w_3 , w_3+ 2(\delta(v)-3)\Delta w_8\} \geq w_3\!+\!10\Delta w_8$,
which is larger than any of  $(16+2k_8)\Delta w_8$ and $36 \Delta w_8$ by \refe{8new-1}.
}\medskip

\medskip
\noindent
Step~5:
In Step~5, the algorithm invokes ${\tt mis}7(G)$.
Analogously with the previous section,
we  include  the following five constraints into the current set of recurrences.
\eqn{MIS8_solution}{\begin{array}{*{20}l}
 C(\mu)\!\leq \! 1.19698^{\frac{w^{\langle 7 \rangle}_3}{w_3}\mu}  ,
  C(\mu)\!\leq \! 1.19698^{\frac{w^{\langle 7 \rangle}_4}{w_4}\mu}  ,
 C(\mu)\!\leq \! 1.19698^{\frac{w^{\langle 7 \rangle}_5}{w_5}\mu}  ,\\
C(\mu)\!\leq \! 1.19698^{\frac{w^{\langle 7 \rangle}_6}{w_6}\mu}  , \mbox{and}~~~
C(\mu)\!\leq \! 1.19698^{\frac{w^{\langle 7 \rangle}_7}{w_7}\mu}  ,
\end{array}}
where $w^{\langle 7 \rangle}_3=0.65077$,
$w^{\langle 7 \rangle}_4=0.78229$, $w^{\langle 7 \rangle}_5=0.89060$, $w^{\langle 7 \rangle}_6=0.96384$ and $w^{\langle 7 \rangle}_7=1$.

\invis
{
  c(1)= 1.19698^(1/x(7)) -x(1); 
  c(2)= 1.19698^(0.96384/x(6)) -x(1); 
  c(3)= 1.19698^(0.89060/x(5)) -x(1); 
  c(4)= 1.19698^(0.78229/x(4)) -x(1); 
  c(5)= 1.19698^(0.65077/x(3)) -x(1); 
}
\medskip

After solving the quasiconvex program, we get
an upper bound $1.19951 $ on the branching factor for all recurrences
 by setting vertex weight:
\eqn{8weight_setting}{
 w_i
=\left\{ \begin{array}{cl}
 0 & \mbox{for  $i=0,1$ and $2$}\\
 0.65844  & \mbox{for $i=3$}\\
 0.78844  & \mbox{for $i=4$}\\
 0.88027 & \mbox{for $i=5$}\\
 0.95345 & \mbox{for $i=6$}\\
 0.98839 & \mbox{for $i=7$}\\
 1  & \mbox{for $i=8$}\\
 w_{8} + (i-8)(w_8\!-\!w_7) & \mbox{for $i\geq 9$.}
  \end{array}
\right.}
This verifies \refl{branch-general} with $\theta=8$.


\section{Proof of \refl{general-optimal}}\label{sec_proof}

We prove \refl{general-optimal} by revealing some structural properties
of  graphs of maximum degree 6, 7 and 8.
Recall that, for a vertex $v$,  $f_v$ denotes the number of edges between $N(v)$ and $N_2(v)$,
and $e_v$ denotes the number of edges in the  graph $G[N(v)]$.
For each neighbor $u\in N(v)$, the outer-degree (resp., inner-degree)
of $u$ at $v$ is $|N(u)\cap N_2(v)|$ (resp.,  $|N(u)\cap N(v)|$).

\subsection{Graphs of maximum degree 6}

The existence of optimal vertices in a reduced graph $G$
of maximum degree 6 without short edges follows from \refl{6optimal}.
When  there is not short edge in a reduced graph with maximum degree 6,
we  see that for each degree-6 vertex $v$ in $G$,
 the inner-degree of any vertex in $N(v)$  at $v$  is at most 2.
Such a graph can have the following types of vertices.

\lem{6optimal}{Let $G$ be a graph of maximum degree 6 and minimum degree $\geq 3$
such that for every degree-$6$ vertex $v$, the inner-degree of each neighbor $u\in N(v)$ at $v$
is at most 2.
If $G$ is not the line graph of a 4-regular graph, then there is a degree-$6$ vertex $v$  that satisfies one of the following:\\
~~ $k_3\geq 1$ or $k_6\leq 3$; \\
~~ $k_6=4$ and $k_5\leq 1$; \\
~~ $k_6=4$, $k_5=2$ and $f_v+(f_v \! -\! |N_2(v)|)+q_v \geq 17$; \\
~~ $k_6=5$, $k_4=1$ and $f_v+(f_v \! -\! |N_2(v)|)+q_v \geq 18$; \\
~~ $k_6=5$, $k_5=1$ and $f_v+(f_v \! -\! |N_2(v)|)+q_v \geq 19$; and \\
~~ $k_6=6$ and $f_v+(f_v \! -\! |N_2(v)|)+q_v \geq 22$.
}

\pf{
Observe that $f_v$ is the sum of the out-degree of neighbors of $v$
and $$e_v\leq 6$$ holds, since the inner-degree of each neighbor  at $v$
  is at most 2.
We assume that $G$ has no  vertex that satisfies one of ``$k_3\geq 1$  or $k_6\leq 3$'' and
``$k_6=4$ and $k_5\leq 1$.'' We consider several cases.

\smallskip\noindent
Case~1.  There is a degree-6 vertex $v$ with  $k_6=5$ and $k_4=1$:
Assume that $f_v+(f_v \! -\! |N_2(v)|)+q_v \leq 16$.
Since  $f_v\geq 16$ by  $e_v\leq 6$,  we see that $f_v \! -\! |N_2(v)|=q_v=0$ and
the degree-4 neighbor $u$ of $v$ is adjacent to a degree-6 vertex
$z\in N(u)\cap N_2(v)$.
Now $N(z) - \{u\}$ contains only degree-6 vertices,
since $z$ has already one degree-4 neighbor.
 Note that $z$ is not adjacent to any vertex in $N[v]-\{u\}$, and
then the outer-degree of vertex $u$ at $z$ is three.
This implies that
$f_{z}\geq 3\times 5+3=18$ and $v$ is a vertex satisfying the condition in the lemma.

In what follows, we further assume that there is no degree-6 vertex  with $k_6=5$ and $k_4=1$.
We choose a degree-6 vertex $v$ with minimum $e_v$ such that $k_6<6$ (if possible) and
then the maximum component in $G[N(v)]$ is maximized.

\smallskip\noindent
Case~2.   $e_v\leq 4$: In this case, we are done because
 $f_v\geq 16+2\times (6-e_v)=20$ for $k_6=4$ and $k_5=2$;
$f_v\geq 17+4=21$ for $k_6=5$ and $k_5=1$; and
$f_v\geq 18+4=22$ for $k_6=6$.

\smallskip\noindent
Case~3.   $e_v=5$: In this case,
 $f_v\geq 16+2\times (6-e_v)=18$ for $k_6=4$ and $k_5=2$;
$f_v\geq 17+2=19$ for $k_6=5$ and $k_5=1$; and
$f_v\geq 18+2=20$ for $k_6=6$.
We only need to consider the case of $k_6=6$,
and assume that $f_v+(f_v \! -\! |N_2(v)|)+q_v \leq 21$, where
we have $f_v \! -\! |N_2(v)|+q_v\leq 1$ by $f_v\geq 20$.
Observe that  $G[N(v)]$ is either a path of length 5 or
 a disjoint union of  a path of $i$ and a cycle of length $5-i$ ($i=0,1,2)$.
We distinguish two subcases.

\smallskip\noindent
(i) $G[N(v)]$ contains a path of length four $u_1u_2u_3u_4u_5$
 (possibly a cycle of length 4 with $u_1=u_5$):
We show that $u_3$ satisfies the lemma.
By $f_v \! -\! |N_2(v)|+q_v\leq 1$,
at most one of $u_2$ and $u_4$ can be adjacent to a vertex in $N(u_3)\cap N_2(v)$;
i.e., the inner-degree of $u_i$ $(i=2,4)$ at $u_3$ is at most 1,
implying that $e_{u_3}\leq 5$ and $u_3$ has only degree-6 neighbors
by our choice of $v$.
Note that $f_{u_3} -|N_2(u_3)|\geq 2$, since $u_1$ and $u_5$ are
common neighbors of two neighbors in $N(u_3)$.
This means that
$f_{u_3}+(f_{u_3}-|N_2(u_3)|)\geq 20+2=22$.

\smallskip\noindent
(ii)  $G[N(v)]$ consists of a path $u_1u_2u_3$ and a triangle:
By $f_v \! -\! |N_2(v)| \leq 1$, we see that for $i=1$ or 3 (say $i=1$),
there is no edge between
$N(u_i)\cap N_2(v)$ and $N(v) -  \{u_i\}$.
In this case, the inner-degree of each of $v$ and $u_2$ at $u_1$ is 1,
 implying that $e_{u_1}\leq 5$ (hence $e_{u_1}=5$).
Hence $u_1$ has only degree-6 neighbors by our choice of $v$.
However, in this case, $G[N(u_1)]$ must be a union of single edge $vu_2$
and   a cycle  of length 4.
Then $u_1$ satisfies the condition of (i).

\smallskip\noindent
Case~4.   $e_v=6$.
In this case,
 $f_v\geq 16$ for $k_6=4$ and $k_5=2$;
$f_v\geq 17$ for $k_6=5$ and $k_5=1$; and
$f_v\geq 18$ for $k_6=6$.

We first consider the case of $k_6=4$ and $k_5=2$.
Assume $f_v+(f_v \! -\! |N_2(v)|)+q_v \leq 16$ (otherwise  we are done), which
implies that $f_v \! -\! |N_2(v)| =q_v=0$.
Let $u\in N(v)$ be a degree-5 neighbor of $v$, and $z\in N(u)\cap N_2(v)$ be
a neighbor of $u$ not in $N[v]$,
where $\delta(z)=6$ (by $q_v=0$) and $z$ is not adjacent to any other
vertices in $N(v) -  \{u\}$ (by  $f_v \! -\! |N_2(v)| = 0$).
Hence the inner-degree of $u$ at $z$ is at most 1, and
$e_z\leq 5$, contradicting our choice of vertex $v$.

Assume that $v$ satisfies ``$k_6=5$ and $k_5=1$'' or ``$k_6=6$.''
Now $G[N(v)]$ with $e_v=6$ is either a cycle of length 6 or
a disjoint union of two triangles.
We distinguish two subcases.

\smallskip\noindent
(i)   $G[N(v)]$  is a cycle of length six $u_1u_2u_3u_4u_5u_6$,
where $u_1$ is assumed to be of degree 5 if $k_5=1$:
By our choice of $v$, it also holds $e_{u_2}=6$, implying that
each of $u_1$ and $u_3$ is adjacent to a vertex in $N(u_2)\cap N_2(v)$.
Analogously
each of $u_4$ and $u_6$ is adjacent to a vertex in $N(u_5)\cap N_2(v)$.
Hence we have $f_v \! -\! |N_2(v)|\geq 4$ and
$f_v+(f_v \! -\! |N_2(v)|)$ is at least $16+4=20$ for ``$k_6=5$ and $k_5=1$''
and $18+4=22$ for $k_6=6$, as required.

\smallskip\noindent
(ii)   $G[N(v)]$ is a disjoint union of two triangles:
In this case, $v$ satisfies ``$k_6=5$ and $k_5=1$'' only, because otherwise
by our choice $G$ would be a 6-regular graph such that for each vertex $v'$ in it,
$G[N(v)]$ is a disjoint union of two triangles and then $G$ would be the line graph of a 4-regular graph.
Assume that $f_v+(f_v \! -\! |N_2(v)|)+q_v \leq 18$ (otherwise we are done),
where $(f_v \! -\! |N_2(v)|)+q_v\leq 1$.
Let $u\in N(v)$ be the degree-5 neighbor of $v$, and
let $\{z_1,z_2\}=N(u)\cap N_2(v)$.
By $ q_v\leq 1$, one of $z_1$ and $z_2$ (say $z_1$) is of degree 6.
Since $e_{z_1}=6$ by our choice of $v$, the inner-degree of $u$ at $z_1$ is 2,
and hence $z_1$ is adjacent to a neighbor in $N(v)\cap N(u)$.
Now $f_v \! -\! |N_2(v)|=1$ and hence $q_v=0$. The vertex $z_2$ also needs to be
adjacent to a neighbor in $N(v)\cap N(u)$,
indicating $f_v \! -\! |N_2(v)|\geq 2$, a contradiction.
}

\subsection{Graphs of maximum degree 7}
To prove the existence of optimal vertices in graphs of maximum degree 7,
we investigate the structure of 7-regular graphs.

A vertex $v$  is called a \emph{$(j,k)$-clique-type} vertex
 if the induced graph $G[N(v)]$ is a disjoint union of a $j$-clique and a $k$-clique, where $j+k=\delta(v)$.

\lem{7regular}{Let $G$ be a 7-regular graph such that the inner-degree of each neighbor $u\in N(v)$
of every vertex $v$ is at most 3.
If  $G$ is not the line graph of a (4,5)-bipartite graph, then
 it has a vertex $v$ such that $f_v+(f_v \! -\! |N_2(v)|)\geq 26$.
}

\pf{Now $f_v=42 -2e_v$ is an even number.
Assume that  for every vertex $v$, $f_v+(f_v \! -\! |N_2(v)|)\leq 25$ otherwise we are done.
Before we prove that $G$  is the line graph of a (4,5)-bipartite graph,
we first show four properties (P0), (P1), (P2) and (P3) on a vertex $v$ in $G$.

 \smallskip \noindent
 (P0) {\em It holds that $f_v\in \{22, 24\}$ and $e_v\geq 9$.}

Now the inner-degree of  each neighbor $u\in N(v)$ at $v$ is at most 3,
i.e., the outer-degree of $u$ of  $v$ is at least 3
in a 7-regular graph, which implies
 $f_v\geq 3\times 7=21$ and hence $f_v\geq 22$ by parity.
By $f_v=42 -2e_v$, we see that $f_v+(f_v \! -\! |N_2(v)|)\leq 25$
implies   $f_v\leq 24$ and   $e_v\geq 9$.

\smallskip \noindent
(P1) {\em If a vertex $v$  is not $(3,4)$-clique-type, then there is no 4-clique  in $G[N(v)]$.}

If $G[X]$ is a 4-clique for some subset $X\subseteq N(v)$, then
the inner-degree of a neighbor $u\in X$ is already 3 and
the remaining $e_v-6\geq 9-6=3$ edges must form a triangle in  $N(v) -  X$,
indicating that  $v$ would be (3,4)-clique-type.

\smallskip \noindent
(P2) {\em If there is no 4-clique  in $G[N(v)]$, then there are
four neighbors $u_{1},u_{2},u_{3},u_{4} \in N(v)$ such that
there is at least one edge  between $N(u_i)\cap N(v)$ and $N(u_i)\cap N_2(v)$.}

Since $f_v\leq 24$ by (P0), there are at least four neighbors $u_{1},u_{2},u_{3},u_{4}\in N(v)$,
each of which has outer-degree 3 at $v$ (i.e., $|N(u_i) -   N_2(v)|=4$).
If there is no edge between $N(u_i)\cap N(v)$ and $N(u_i)\cap N_2(v)$,
then     $e_{u_i}\geq 9$ by  (P0) implies that $N(u_i)\cap N(v)$ and $N(u_i) -   N_2(v)$
induce a 3-clique and 4-clique, respectively
(where $N(u_i)\cap N(v)$ induces a 3-clique),
which contradicts that   $\{u_i\}\cup (N(u_i)\cap N(v))$ does not induce a 4-clique.

\smallskip \noindent
(P3) {\em If there is exactly one edge between  $X$ and $X -  N(v)$
for some set $X$ of four vertices in $N(v)$,
then $v$ is not $(3,4)$-clique-type and $f_v\geq 24$. }

If $f_v\leq 22$ (i.e., $e_v\geq 10$) then $G[X]$ needs to be a 4-clique and
the inner-degree of some vertex  in $X$ at $v$ would be 4.

We are ready to prove the lemma by using (P0),   (P1), (P2)  and (P3).
If the graph is not the line graph of a (4,5)-bipartite graph,
then we can always choose  a vertex $v$  that is not $(3,4)$-clique-type so that  $f_v\in \{22,24\}$ is maximized.
By (P1) and (P2),  there are  four neighbors $u_{i}\in N(v)$, $i=1,2,3,4$ such that
there is at least one edge  between $N(u_i)\cap N(v)$ and $N(u_i)\cap N_2(v)$.

(i)  For each $i$, there are two such edges: Then
we get $f_v \! -\! |N_2(v)|\geq 4$ and it would hold $f_v+(f_v \! -\! |N_2(v)|)\geq 22+4=26$,  a contradiction.

(ii) For some $i$, there is exactly one such edge: In this case,
we get $f_v \! -\! |N_2(v)|\geq 2$.
Since $u_i$ satisfies the condition of (P3), it holds $f_{u_i}\geq 24$, and
by the choice of $v$, we have $f_v\geq f_{u_i}\geq 24$.
Hence it would hold $f_v+(f_v \! -\! |N_2(v)|)\geq 24+2=26$, a contradiction.
}

\lem{7optimal}{Let $G$ be a reduced graph  of maximum degree 7. Assume that $G$ has no short edges.
Then $G$ has at least one optimal vertex.}

\pf{We assume that every degree-7 vertex $v$ has at least six neighbors of degree 7 and
satisfies $N^*(v)= \emptyset$,
otherwise the lemma holds.
Each neighbor of $v$ is adjacent to at least two vertices in $N_2(v)$ (since $N^*(v)= \emptyset$)
and
each degree-7 neighbor of $v$ is adjacent to at least three vertices in $N_2(v)$ (since here is no short edge).
Hence $f_v\geq 20$. If  $v$ is adjacent to a degree-3 vertex, then $v$ is an optimal vertex by the definition of optimal vertices.
If the graph is a 7-regular graph, then there
is an optimal vertex by \refl{7regular}.
Otherwise, we can always find a vertex $v$ with $k_7=6$ and $k_3=0$.

Let $u_1$ be the neighbor of $v$ such that $4\leq \delta(u_1)\leq 6$. If $u_1$ is not adjacent to any other vertex in $N(v)$, then $f_v\geq \delta(u_1)-1+3\times 6 =\delta(u_1)+17$ and $v$ would be an optimal vertex. Otherwise, $u_1$ is adjacent to a degree-7 vertex $u_2\in N(v)$. If $u_2$ has outer-degree at least 5 at $v$, then $f_v\geq 5+3\times 5+2=22$ and $v$ would be an optimal vertex. We can assume that $u_2$ has outer-degree  3 or 4 at $v$. If there are at least two edges between $N(u_2)\cap N(v)$ and $N(u_2)\cap N_2(v)$, then $f_v \! -\! |N_2(v)|\geq 2$ and $v$ would be an optimal vertex. Otherwise, there is at most one edge between $N(u_2)\cap N(v)$ and $N(u_2)\cap N_2(v)$. Since $\{|N(u_2)\cap N(v)|, |N(u_2)\cap N_2(v)|\}=\{3,4\}$, then $e_{u_2}\leq 10$ and $f_{u_2}\geq22$. Note that $u_2$ is also adjacent to
a vertex $u_1$ with degree $<7$. Then $u_2$ will be an optimal vertex for this case.
}

\subsection{Graphs of maximum degree 8}

To prove the existence of optimal vertices in graphs of maximum degree 8,
we investigate the structure of 8-regular graphs.

First of all, we consider 8-regular graph such that the inner-degree of each neighbor $u\in N(v)$
of every vertex $v$ is at most 3.
Now $f_v=56 -2e_v$ is an even number.
 This properties will be used in the following lemmas several times.
A vertex $v$ is called {\em bridge-type} if $G[N(v)]$
contains a bridge $u_1 u_2$ between $X\subseteq N(v)$ and $N(v) -  X$ such that
(i) $|X|=|N(v) -  X|$; or
(ii) the inner-degree of each $u_i$ at $v$ is 3.

\lem{bridge}{Let $G$ be a 8-regular graph such that the inner-degree of each neighbor $u\in N(v)$
of every vertex $v$ is at most 3.
If $G$ contains a bridge-type vertex, then
there is a vertex $v$  such that $f_v+(f_v \! -\! |N_2(v)|)\geq 36$.
}

\pf{
Let $v$ a bridge-type vertex $v$, and $u_1u_2$ be the bridge of $v$
 between $X_1$ and $X_2=N(v) -  X$, where
$u_i\in X_i$ and $|X_1|\leq |X_2|$
without loss of generality.

We first prove that $f_v\geq 34$.
Note that $|X_1|\geq 3$, since otherwise the inner-degree of $u_1$ at $v$ would be at most
$|X_1|-1\leq 2$.
When $|X_1|=3$, the inner-degree  of each of the two vertices in $X_1 -  \{u_1\}$
is at most 2 at $v$, since $u_1u_2$ is a bridge in $G[N(v)]$.
When $|X_1|=|X_2|=4$, $G[X_i]$ contains at least one pair of
non-adjacent vertices, since otherwise the  inner-degree of $u_i$ would be
four, and $X_i$ contains at least one vertex whose inner-degree at $v$ is at most 2.
In any case of $|X_1|\in \{3,4\}$,
 $N(v)$ contains at least two neighbors whose outer-degree at $v$ is at least 5.
This implies that $f_v\geq 6\times4+2\times5=34$.

We further assume that $N(v)$ consists of two (resp., six)
neighbors whose inner-degree at $v$ are 2 (resp., 3), since otherwise
$f_v\geq =35$ (hence $f_v\geq 36$ by parity) and we are done.
Note that there are at least three neighbors in $N(v)$
whose inner-degree are 2 if $|X_1|=3$ (2 in $X_1$ and 1 in $X_2$).
So it holds $|X_1|=4$ and
the inner-degree of $u_i$ at $v$ is 3.

The outer-degree of  $u_i$ ($i=1,2$) at $v$ is 4
(i.e., $|N(u_i)\cap N[v]|=4$),
and the induced graph $G[N(u_i)\cap N[v]]$ contains at least two pairs of
non-adjacent vertices (since there is only one edge between $N[u_i]\cap N(v)$ and $N(v)-N[u_i]\cap N(v)$).
If there is no edge between $N(u_i)\cap N[v]$ and $N(u_i)\cap N_2(v)$, then $e_{u_i}\leq10$ and $f_{u_i}\geq 56-2\times10=36$.
Hence we can assume that $u_i$ has a common neighbor in $N_2(v)$ with a vertex in $N(v) - \{u_i\}$.

Let $u'_i$ ($i=1,2$) be the two neighbors in $N(v)$
whose inner-degree at $v$ is 2.
If there is no edge between   $N(u'_i)\cap N[v]$ and $N(u_i)\cap N_2(v)$,
then each vertex in $N(u'_i)\cap N[v]$ has outer-degree 5 at $u'_i$.
We get that $f_{u'_i}\geq 3\times5+5\times4=35$
(hence $f_{u'_i}\geq 36$ by parity).
Hence we can assume that $u'_i$ has a common neighbor in $N_2(v)$ with a vertex in $N(v) - \{u'_i\}$.

Thus there are four vertices $\{u_1,u_2,u'_1,u'_2\}$ each of who has a common neighbor in $N_2(v)$ with another vertex in
$N(v)$, which implies that $f_v \! -\! |N_2(v)|\geq 4/2=2$, and
we have that $f_v+(f_v \! -\! |N_2(v)|)\geq 34+2=36$.
}
\bigskip

Recall that a vertex $v$  is $(j,k)$-clique-type if the induced graph $G[N(v)]$ is a disjoint union of a $j$-clique and a $k$-clique, where $j+k=\delta(v)$.
A vertex $v$ is called {\em semi-clique-type} if a set $X\subseteq N(v)$ induces a 4-clique,
there is no edge between $X$ and $N(v) -  X$,
and  $X\subseteq N(v)$ contain at least one pair of non-adjacent vertices.
Hence a semi-clique-type vertex is not $(4,4)$-clique-type.

\lem{semi}{Let $G$ be a 8-regular graph such that the inner-degree of each neighbor $u\in N(v)$
of every vertex $v$ is at most 3.
 If $G$ contains a semi-clique-type vertex, then
there is a vertex $v$  such that $f_v+(f_v \! -\! |N_2(v)|)\geq 36$.
}

\pf{Let $v$ be a semi-clique-type vertex, and assume that $X\subseteq N(v)$ induce a 4-clique.
Clearly $f_v\geq 56-2\times11=34$ (since $e_v\leq 11$).
Assume that $N(v) -  X$ contains only one pair of non-adjacent vertices $u_1$ and $u_2$,
since otherwise we obtain $f_v\geq 56-2\times10=36$.
Let $N(v) -  X=\{u_1,u_2,u_3,u_4\}$.
If each of $u_3$ and $u_4$ is adjacent to a vertex in $N(u_1) -  N[v]$,
then we have  $f_v -|N_2(v)|\geq 2$, and $f_v+(f_v \! -\! |N_2(v)|)\geq 34+2=36$.
If none of $u_3$ and $u_4$ is adjacent to any vertex in $N(u_1) -  N[v]$,
then all of $v$, $u_3$ and $u_4$ have inner-degree 2 at $u_1$, and
this implies that $f_{u_1}\geq 36$.
Finally if exactly one of $u_3$ and $u_4$ is adjacent to a vertex in $N(u_1) -  N[v]$, then
 $u_1$ is bridge-type, and $f_{u_1}+(f_{u_1}-|N_2({u_1})|)\geq  36$  by \refl{bridge}.
}

\lem{deg8-36}{Let $G$ be a 8-regular graph such that the inner-degree of each neighbor $u\in N(v)$
of every vertex $v$ is at most 3.
Assume that $G$ is not the line graph of a 5-regular graph.
Then  $G$ has a vertex $v$ such that $f_v+(f_v \! -\! |N_2(v)|)\geq 36$.
}

\pf{
By \refl{bridge} and \refl{semi}, it suffices to show that
 there is a desired vertex or a bridge-type or semi-clique-type vertex.
Let $v$ be a vertex that is not $(4,4)$-clique-type. This vertex always exists since the graph is not
the line graph of a 5-regular graph.
Assume that $v$ is not bridge-type or semi-clique-type.
 Hence no four neighbors of $v$ induce a 4-clique.

Assume $f_v\leq 34$, otherwise we are done.
This implies that there are at least six neighbors $u_i\in N(v)$, $i=1,\ldots,6$
each of which has outer-degree 4 at $v$ (i.e., $|N[u_i]\cap N(v)|=4$).

 If for each $i$, there are at least two edges between $N(u_i)\cap N(v)$ and $N(u_i)\cap N_2(v)$, then
we have  $f_v -|N_2(v)|\geq 12/2=6$, and $f_v+(f_v \! -\! |N_2(v)|)\geq 32+6> 36$.
Assume that for some $i$,
there is at most one edge  between $N(u_i)\cap N(v)$ and $N(u_i)\cap N_2(v)$.
If there is exactly one edge  between $N(u_i)\cap N(v)$ and $N(u_i)\cap N_2(v)$, then
$u_i$ is bridge-type.
Assume that
there is no edge  between $N(u_i)\cap N(v)$ and $N(u_i)\cap N_2(v)$.
Recall that no four vertices in $N(v)$ induce a 4-clique.
Hence $N(u_i)\cap N[v]$ is not a 4-clique either, and this means that $u_i$ is semi-clique-type.
}

\lem{8optimal}{Let $G$ be a reduced graph of maximum degree 8.
If $G$ has no short edges, then $G$ has at least one optimal vertex.}

\pf{
This lemma follows from the definition of optimal vertices and \refl{deg8-36} directly.
If the graph $G$ is not a 8-regular graph,
we can always find a degree-8 vertex such that $k_8<8$, which is an optimal vertex by the definition.
Otherwise,  $G$ is a 8-regular graph.
Since  $G$ has no short edges,  the inner-degree of each neighbor $u\in N(v)$
of every vertex $v$ is at most 3.
Then by \refl{deg8-36} there is either an optimal vertex or the graph is the line graph of a 5-regular graph.
However, the later case is impossible, since the line graph of a 5-regular graph
must have been reduced by the reduction rules.
}

\section{Concluding Remarks}
Before the measure-and-conquer method was developed, most fast algorithms for
the maximum independent set problem consisted of a large number of branching rules,
which may make the algorithms impractical and hard to analyze.
The measure-and-conquer method  allows us to design simple algorithms
for the maximum independent set problem probably with an aid of sophisticated analysis.
With this method, we get the recurrence \refe{max-deg} for branching on
a vertex $v$ of maximum degree $d$, which usually becomes the worst case of  algorithms to any
MIS-$\theta$ $(3\leq \theta\leq 8)$.
To analyze \refe{max-deg}, we need to do both  of
 (i) checking all possible neighbor-degree  $(k_3, k_4,\dots,k_d)$
of the neighbors of $v$; and (ii) deriving lower bounds on the term $\Delta (\overline{N[v]})$.

For (i),  the previous papers either to try to reduce the number of cases to be checked by a relaxed argument
(and then get  worse recurrences) or list up a huge number of recurrences for all possible combinations
(which may not be easy to check by hand).
In this paper, we devised a new lemma (\refl{simplify_analysis}) that can reduce the number of cases to a quite small number
without losing the optimality of branching factors.
 In the branch-and-reduce paradigm, this is useful to simplify  analysis of algorithms
and  can make a design process of fast algorithms much easier.

For (ii), there are may techniques used to derive good bounds on $\Delta (\overline{N[v]})$.
With the reduction rule by domination,
Fomin \emph{et al.} ~\cite{Fomin:is} got that $\Delta (\overline{N[v]})\geq d \Delta w_d$.
With  branching on vertices with satellites (which is extended to unconfined vertices later in \cite{XN:3MIS}),
Kneis \emph{et al.}~\cite{kneis:MIS} showed 
 $\Delta (\overline{N[v]}) \geq 2d\Delta w_d$ in the worst case $k_d=d$ of \refe{max-deg}
(this is also used in Bourgeois \emph{et al.}'s algorithm~\cite{Bourgeois:alg}).
In this paper, by using the new branching rule to short edges, for the worst case  of \refe{max-deg}, we
improved the bounds on $\Delta (\overline{N[v]})$ to $3d\Delta w_d$ for $d=6,7$ and to $4d\Delta w_d$ for $d\geq 8$, respectively.
By choosing an optimal vertex whose existence is ensured by a graph theoretical argument,
 we  further increased the bound on $\Delta (\overline{N[v]})$ to $(4d+4)\Delta w_d$ for the case of ``$d= 8$ and $k_d=d$,''
which is the final worst case in our algorithms. Branching on a degree-8 vertex $v$ with eight degree-8 neighbors  and 36 edges between $N(v)$ and $N_2(v)$ (i.e., $f_v=36$ for  $k_8=8$)  is one of the crucial bottlenecks in our algorithm for MIS now.

\section*{Acknowledgements}
A preliminary version of this paper with a result of $1.2002^nn^{O(1)}$ was presented in
the 24th international symposium on algorithms and computation (ISAAC 2013) and appeared as~\cite{xn:mis}.
In the full version, we further reduce several bottlenecks by refining the definition of ``optimal vertices'' and then finally break the barrier of 1.2 in the base of the running time.

\end{document}